\documentclass[aps,prd,amsmath
 ,eqsecnum            
 ,preprintnumbers
 ,nofootinbib         
 ,twocolumn         
 ]{revtex4}
 \usepackage[dvips]{graphicx}
 \usepackage{amsmath}
 \usepackage{amsfonts}
 \usepackage{amssymb}
 \usepackage{longtable}   

 \allowdisplaybreaks[4]     

 \newcommand{\eqlabel}[1]{
   \label{#1}
 }
 \newcommand{\seclabel}[1]{
   \label{#1}
 }
 \newcommand{\figlabel}[1]{
   \label{#1}
 }
 
 \newcommand{\LT}{Lema\^{\i}tre-Tolman}
 
 \newcommand{\td}[2]{\frac{{\rm d} {#1}}{{\rm d} {#2}}}
 \newcommand{\tdn}[3]{\frac{{\rm d}^{#1} {#2}}{{\rm d} {#3}^{#1}}}
 
 \newcommand{\pd}[2]{\frac{\partial {#1}}{\partial {#2}}}

 \newcommand{\nn}{\nonumber}

 \newcommand{\Chr}[3]{\mbox{\small\( \begin{Bmatrix}#1\\#2#3\end{Bmatrix}\)}}
 
 \newcommand{\df}{\ {\overset {\rm def} =}\ }
 \newcommand{\dr}[2]{\frac {{\rm d} {#1}} {{\rm d} {#2}}}

 \newcommand{\llim}[1] {\ {\underset {#1} {\longrightarrow}}\ }
 
 \renewcommand{\o}[1] {\overline{#1}} 

\begin{document}
 \sffamily

 \title{
 Physical and Geometrical Interpretation of the $\epsilon \leq 0$ Szekeres Models.
 }

 \author{Charles Hellaby}
 \affiliation{Department of Mathematics and Applied Mathematics, \\
 University of Cape Town, Rondebosch 7701, South Africa}
 \email{cwh@maths.uct.ac.za}

 \author{Andrzej Krasi\'{n}ski}
 \affiliation{N. Copernicus Astronomical Center, Polish Academy of Sciences, \\
 Bartycka 18, 00 716 Warszawa, Poland}
 \email{akr@camk.edu.pl}

 \date{}

 \begin{abstract}
   We study the properties and behaviour of the quasi-pseudospherical
and quasi-planar Szekeres models, obtain the regularity conditions, and analyse
their consequences.  The quantities associated with ``radius" and ``mass" in the
quasi-spherical case must be understood in a different way for these cases.  The
models with pseudospherical foliation can have spatial maxima
and minima, but no origins.  The ``mass" and ``radius" functions may be one
increasing and one decreasing without causing shell crossings. This case most
naturally describes a snake-like, variable density void in a more gently varying
inhomogeneous background, although regions that develop an overdensity are also
possible.  The Szekeres models with plane foliation can have neither spatial
extrema nor origins, cannot be spatially flat, and they cannot have more
inhomogeneity than the corresponding Ellis model, but a planar surface can be
the boundary between regions of spherical and pseudospherical foliation.
 \end{abstract}

 \maketitle

 \begin{center}

 \end{center}

 \section{Introduction}
 \seclabel{Intro}

The Szekeres metric is important because, as a model with 5 arbitrary functions,
it exhibits features of non-linear gravitation that less general models cannot.
It is an exact inhomogeneous solution of the Einstein field equations (EFEs)
that has a realistic equation of state (dust) suitable for the post
recombination universe, it has no Killing vectors.  It is necessary to pay more
attention to models with little symmetry in order to better understand all the
features and possibilities of General Relativity, and therefore to better model
the structures of our universe.

Although there have been a number of papers that investigate the Szekeres metric
generally \cite{Szek1975a, BoTo1976, GoWa1982a, GoWa1982b}, and several papers
that investigate the quasi-spherical case in particular \cite{Szek1975b,
Bonn1976a, BoST1977, Bonn1976b, DeSo1985, Bonn1986, BoPu1987, Szek1980}, there
have been none that specifically look at the quasi-pseudospherical
and quasi-planar cases.  This is probably because we have a good
understanding of spherical gravity from Newtonian theory, and so relativistic
analyses of spherically symmetric metrics were easily developed.  Without
spherical symmetry, or a slight variation of it, familiar relationships, such as
that between the mass inside a sphere and the gravitational potential, do not
apply, so it is much more difficult to interpret the equations physically.

We here set out to improve our understanding of the quasi-planar and
quasi-pseudospherical models, and thus enhance their
usability, by analysing their physical and geometric properties.  The main
challenge is to develop a re-interpretation of quantities such as ``radius" and
``mass" that cannot retain the meaning they have in (nearly) spherical models.

The appearance of the first paper \cite{Bole2006} to produce an explicit model
using the quasi-spherical Szekeres metric, that of a void adjacent to a cluster,
and to plot the evolution of its density, is an encouraging development.  If the
other Szekeres cases are sufficiently well understood, explicit models can be
produced from these too.

   Our methods below are to (a) analyse how the metric functions affect the
geometry, the matter distribution, and the evolution, (b) derive regularity
conditions on the metric for well behaved matter, curvature and evolution, (c)
compare with other metrics that have planar and pseudospherical
symmetry, and (d) produce one or two simple examples.

 \section{The Szekeres Metric}
 \seclabel{SzekMetric}

In this section we will present the metric and its basic relationships, but we
will refrain from any physical interpretation, reserving that for a later
section.  Once all the features and properties of the model are established, we
will collect the results, discuss the meaning of the various functions, and
attempt an interpretation of the model.

   Our notation is that of \cite{HeKr2002}, for which this is a follow up.  The
LT-type Szekeres metric \cite{Szek1975a, Szek1975b, Kras1997, PlKr2006}%
 \footnote{In Ref. \cite{PlKr2006} this family of the Szekeres solutions is called
the $\beta' \neq 0$ family.}
 is:
 \begin{align}
  ds^2 = - dt^2 + \frac{(R' - R \frac{\textstyle E'}{\textstyle E})^2}
                       {(\epsilon + f)} dr^2
            + R^2 \frac{(dp^2 + dq^2)}{E^2} ~,   \eqlabel{2.1}
 \end{align}
 where ${}' \equiv \partial/\partial r$, $\epsilon = \pm1,0$ and $f = f(r) \geq
-\epsilon$ is an arbitrary function of $r$.  The function $E$ is given by
 \begin{align}
   E(r,p,q) & = \frac{S}{2} \left\{ \left( \frac{p - P}{S} \right) ^2
      + \left( \frac{q - Q}{S} \right) ^2 + \epsilon \right\} ~, \eqlabel{Edef} \\
      \epsilon & = 0, \pm 1 ~, \nonumber
 \end{align}
where $S = S(r)$, $P = P(r)$, and $Q = Q(r)$ are arbitrary functions. In the
original parametrisation of Szekeres, $E$ had the form
 \begin{align}\eqlabel{Edeforig}
   E(r,p,q) = A (p^2 + q^2) + 2 B_1 p + 2 B_2 q + C ~,
 \end{align}
 where%
 \footnote{
In the original parametrisation of Szekeres, the $\epsilon$ is an arbitrary
function of $r$. If nonzero, this function can be scaled to $+1$ or $-1$ by the
rescalings of the other functions: $R = \sqrt{|\epsilon|} \widetilde{R}$, $E =
\sqrt{|\epsilon|} \widetilde{E}$, $f =|\epsilon| \widetilde{f}$. The scalings
cannot change the signs of $\epsilon$ and of $f$.
 }
 \begin{align}
   A & = \frac{1}{2S} ~,~~ B_1 = \frac{-P}{2S} ~,~~
      B_2 = \frac{-Q}{2S} ~, \nn \\
   C & = \frac{P^2 + Q^2 + \epsilon S^2}{2S} ~,~~
      4(AC - B_1^2 - B_2^2) = \epsilon ~.   \eqlabel{Econd}
 \end{align}
 The function $R = R(t,r)$ satisfies the Friedmann equation for dust
 \begin{align}
   \dot{R}^2 = \frac{2M}{R} + f ,  \eqlabel{2.5}
 \end{align}
where $\dot{{}} \equiv \partial/\partial t$ and $M = M(r)$ is another arbitrary
function of coordinate $r$.  It follows that the acceleration of $R$ is
 \begin{align}
   \ddot{R} = \frac{-M}{R^2}.  \eqlabel{Rddot}
 \end{align}
 Solving (\ref{2.5}), the evolution of $R$ depends on the value of $f$; it
can be: \\
 ${}$~~hyperbolic, $f > 0$:
 \begin{align}
   R & = \frac{M}{f} (\cosh \eta - 1) ~,   \eqlabel{hypevRS}
   \\ \nonumber \\
   (\sinh \eta - \eta) & = \frac{f^{3/2} \sigma (t - a)}{M} ~,
   \eqlabel{hypevtS}
 \end{align}
 ${}$~~parabolic, $f = 0$:
 \begin{align}
   R = \left( \frac{9 M (t - a)^2}{2} \right)^{1/3} ~,   \eqlabel{evoRflat}
 \end{align}
 ${}$~~or elliptic, $f < 0$:
 \begin{align}
   R & = \frac{M}{(-f)} (1 - \cos \eta) ~,   \eqlabel{ellevRS}
   \\ \nonumber \\
   (\eta - \sin \eta) & = \frac{(-f)^{3/2} \sigma (t - a)}{M} ~,
   \eqlabel{ellevtS}
 \end{align}
 where $a = a(r)$ is the last arbitrary function, giving the local time of the
big bang or crunch $R = 0$ and $\sigma = \pm 1$ permits time reversal. More
correctly, the three types of evolution hold for $f/M^{2/3} >,=,< 0$, since $f =
0$ at a spherical type origin for all 3 evolution types.  The behaviour of
$R(t,r)$ is identical to that in the {\LT} (LT) model, and is unaffected by
$(p,q)$ variations.

The 6 arbitrary functions $f$, $M$, $a$, $S$, $P$ and $Q$ give us 5 functions to
control the physical inhomogeneity, plus a choice of the coordinate $r$.
Note, however, that in the case $\epsilon = 0$ we are free to redefine the
functions $R$, $S$, $f$ and $M$ as follows:
 \begin{equation}\eqlabel{2.12}
(R, S, f, M) = (\alpha \widetilde{R}, \widetilde{S} / \alpha , \alpha^2
\widetilde{f}, \alpha^3 \widetilde{M}),
 \end{equation}
 where $\alpha = \alpha(r)$ is an arbitrary function, and the form of the
metric, the density and the evolution equations will not change.  In particular,
we can choose $\alpha$ so that $\widetilde{S} = 1$.

   The density and Kretschmann scalar are functions of all four coordinates
 \begin{align}
   8 \pi \rho & = G_{tt}
      = \frac{2 (M' - 3 M E' / E)}{R^2 (R' - R E' / E)},
         \eqlabel{2.13} \\ \nonumber \\
   {\cal K} & = R^{\alpha \beta \gamma \delta} R_{\alpha \beta \gamma \delta}
      = (8 \pi)^2 \left[ \frac{4}{3} \o{\rho}^2 - \frac{8}{3}
        \o{\rho} \rho + 3 \rho^2 \right],
        \eqlabel{KretschDef}
 \end{align}
 where
 \begin{align}
   8 \pi \o{\rho} = \frac{6 M}{R^3}   \eqlabel{rhobardef}
 \end{align}
 is some kind of mean density.  For all $\rho$ and $\o{\rho}$ we
have ${\cal K} \geq 0$, but assumptions of positive mass and density require
$\rho \geq 0$ and $\o{\rho} \geq 0$.  The flow properties of the comoving matter
were given for any $\epsilon$ value in \cite{HeKr2002}.  For further discussion
of this metric see \cite{Kras1997,PlKr2006}.

   In the following, we will call the comoving surfaces of constant $r$ ``shells'',
and paths that follow constant $p$ \& $q$ will be termed ``radial''.  We will use
the term ``hyperbolic'' to describe the time evolution for $f > 0$, and
``pseudospherical'' or ``hyperboloidal'' to describe the shape of the $(p,q)$
2-surfaces when $\epsilon = -1$.  To make it clear the shells are quite different
from spheres, we will call $r$ the ``p-radius'' or ``h-radius'', $R$ the ``areal
p-radius'' or ``areal h-radius', and $M$ the ``p-mass'' or ``h-mass'', in the
planar or pseudospherical cases, respectively.  However, we will use ``radius''
generically when more than one $\epsilon$ value is considered.

 \subsection{Singularities}
 \seclabel{Sing}

The bang or crunch occur when $t = a$ or $t = 2\pi M/ (- f)^{3/2} + a$, which
makes $R = 0$ and both $\rho$ and $\cal K$ divergent. Shell crossings
happen when surfaces (``shells") of different $r$ values intersect, i.e. $R' = R
E' / E$ and $M' \neq 3 M E' / E$.  Also $\rho$ but not ${\cal K}$ passes through
zero where $E'/E$ exceeds $M'/3M$.

 \subsection{Special Cases and Limits}
 \seclabel{SCaL}

   The {\LT} model is the spherically symmetric special case $\epsilon = +1$,
$E' = 0$.

   The Ellis metrics \cite{Elli1967} result as the special case $E' = 0$; they
are the LT model and its counterparts with plane and pseudospherical symmetry.

   The vacuum case is $(M' - 3 M E' / E) = 0$, which implies
$E' = M' = 0 = S' = P' = Q'$.  For $M \neq 0$ this gives pseudospherical
and planar equivalents of the Schwarzschild metric
\cite{CaDe1968} (see section \ref{VacMatch}).

   The null limit is obtained by taking $f \rightarrow \infty$ after a suitable
tranformation.  In this limit the `dust' particles move at light speed
\cite{Hel96, Bon97} and the metric becomes a pure radiation Robinson-Trautman
metric of Petrov type D (see \cite{SKMHH03} eq (28.71) with (28.73)).

   The Kantowski-Sachs (KS) type Szekeres metric is in fact a regular limit of
the LT type Szekeres metric \cite{Hel96,PlKr2006}.

 \subsection{Basic Physical Restrictions}
 \seclabel{BPR}

 \begin{enumerate}

 \item   In order to keep the metric signature Lorentzian we must have
 \begin{align}
   \epsilon + f \geq 0 ~,
   \eqlabel{epsf>=0}
 \intertext{and in particular}
   \epsilon + f > 0 ~~~~\mbox{and}~~~~ R' - \frac{R E'}{E} \neq 0 ~,
   \eqlabel{epsf>0}
 \end{align}
 while
 \begin{align}
   \epsilon + f = 0 ~~~~\mbox{where}~~~~ R' = \frac{R E'}{E} ~.
   \eqlabel{epsf=0}
 \end{align}
Clearly, pseudospherical foliations, $\epsilon = -1$, require
$f\geq 1$, and so are only possible for regions with hyperbolic evolution, $f >
0$.  Similarly, planar foliations, $\epsilon = 0$, are only possible for regions
with parabolic or hyperbolic evolution, $f \geq 0$; whereas spherical foliations
are possible for all $f \geq -1$.

 \item   We require the metric to be non degenerate \& non singular,
except at the bang or crunch.  For a well behaved $r$ coordinate then, we need
to specify
 \begin{align}
   \infty > \frac{(R' - R E'/E)^2}{(\epsilon + f)} > 0 ~.
   \eqlabel{grrFinite}
 \end{align}
 Whilst failure to satisfy this may only be due to bad coordinates, there should exist
a choice of $r$ coordinate for which it holds.

 \item   The density must be positive, and the Kretschmann scalar must be
finite, i.e.
 \begin{align}
   \infty > \frac{M' - 3 M E' / E}{R' - R E' / E} \geq 0 ~.   \eqlabel{PosDen}
 \end{align}

 \item    We assume
 \begin{align}
   R \geq 0 ~,~~~~ M \geq 0 ~~~~\mbox{and}~~~~ S > 0 ~.
 \end{align}
 The sign of $S$, and hence of $E$ can be flipped without changing the metric, but $S = 0$
is not acceptable.

 \item   The various arbitrary functions should have sufficient continuity
 --- $C^1$ and piecewise $C^3$
 --- except possibly at a spherical origin.

 \end{enumerate}

 \subsection{3-spaces of constant $t$}
 \seclabel{3spaces}

   It is known from \cite{BeEaOl77} that when $\epsilon = +1$ these 3-spaces are
conformally flat, and it is easy to verify, using Maple \cite{Maple} and
GRTensor \cite{GRT} that the Cotton-York tensor is zero for all $\epsilon$ (see
\cite{PlKr2006}, section 19.11, exercise 19.14, and theorem 7.1).

   Calculating the Riemann tensor for the constant $t$ spatial sections of
(\ref{2.1}), we find
 \begin{align}
   {}^3\!R_{rprp} & = {}^3\!R_{rqrq} \nn \\
      & = \frac{- R}{E^2 (\epsilon + f)}
         \left( R' - \frac{R E'}{E} \right)
         \left( \frac{f'}{2} - \frac{f E'}{E} \right)   \eqlabel{2.9} \\
   {}^3\!R_{pqpq} & = \frac{- R^2 f}{E^4}   \eqlabel{2.10} \\
   {}^3\!R & =
   \frac{2 f}{R^2} \left( \frac{2 \left( \frac{f'}{2 f} - \frac{E'}{E} \right)}
         {\left( \frac{R'}{R} - \frac{E'}{E} \right)} + 1 \right) \\
   {}^3{\cal K} & = {}^3\!R^{ijkl} {}^3\!R_{ijkl} =
         \frac{4 f^2}{R^4} \left( \frac{2 \left( \frac{f'}{2 f} - \frac{E'}{E} \right)^2}
         {\left( \frac{R'}{R} - \frac{E'}{E} \right)^2} + 1 \right)
 \end{align}
where equations (2)-(5) of \cite{Hel96} have been used, and the other  curvature
invariants are linearly dependent on these. The flatness condition
${}^3\!R_{abcd} = 0$ requires
 \begin{align}
   & & \epsilon & \neq 0:~~ & & f = 0 &   \eqlabel{3FlatCondepsnot0} \\
   & & \epsilon & = 0:~~ & & R' = E' = f' = f = 0 &   \eqlabel{3FlatCondeps0}
 \end{align}
and the latter is only possible as a limit, or as a Kantowski-Sachs type
Szekeres model \cite{Hel96b}. Interestingly, $(\epsilon + f)$ does not enter any
curvature invariants, and they are all well behaved if $f = 0$. The 2-spaces of
constant $t$ and $r$ have Ricci scalar
 \begin{align}
   {}^2\!R & = \frac{2 \epsilon}{R^2} ~.   \eqlabel{2dRiccipq}
 \end{align}

 \subsection{General properties of $E(r, p, q)$}

   From (\ref{Edef}) we see $E$ has circular symmetry about $p = P$, $q = Q$,
which is a different point for each $r$.  The $E = 0$ locus
 \begin{align}
   (p - P)^2 + (q - Q)^2 = - \epsilon S^2 ~,   \eqlabel{E=0circle}
 \end{align}
 only exists if $\epsilon \leq 0$, and is clearly a circle in the $p$-$q$ plane,
with $E > 0$ on the outside, but becomes a point $p = P$, $q = Q$ if $\epsilon = 0$.
We have
 \begin{align}
   E' & = - \frac{S'}{2} \left\{ \left( \frac{p - P}{S} \right)^2
      + \left( \frac{q - Q}{S} \right)^2 - \epsilon \right\} \nn \\
 &~~~~ - \left( \frac{p - P}{S} \right) P' - \left( \frac{q - Q}{S} \right) Q'
      \eqlabel{E'-pq-h}
 \end{align}
 so the $E' = 0$ locus is also a circle in the $p$-$q$ plane, since it can be
written
 \begin{align}
   \left( \frac{p - P}{S} + \frac{P'}{S'} \right)^2
      + \left( \frac{q - Q}{S} + \frac{Q'}{S'} \right)^2
      = \frac{(P')^2 + (Q')^2}{(S')^2} + \epsilon ~.
      \eqlabel{E'=0circle}
 \end{align}
 With $\epsilon \geq 0$, this locus always exists, and with $\epsilon = -1$ it
only exists if
 \begin{align}
   (S')^2 < (P')^2 + (Q')^2 ~,   \eqlabel{hE'=0exists}
 \end{align}
 with the radius of this circle shrinking to zero as the equality is approached.
Since, if they exist, the distance between the centres of these two circles
never exceeds the sum of their radii
 \begin{align}
   \left|\frac{S}{S'}\right| & \sqrt{(P')^2 + (Q')^2}\; \leq \nn \\
   & \left|\frac{S}{S'}\right| \left(\left|S'\right| \sqrt{-\epsilon}\;
      + \sqrt{(P')^2 + (Q')^2 + \epsilon (S')^2}\; \right)   \eqlabel{Overlap}
 \end{align}
they always intersect, and the intersection points are
 \begin{align}
   \frac{p - P}{S} & = \frac{\epsilon P' S' \pm Q' \sqrt{-\epsilon
      \left\{(P')^2 + (Q')^2 + \epsilon (S')^2\right\}}\;}{(P')^2 + (Q')^2\;} ~, \nn \\
   \frac{q - Q}{S} & = \frac{\epsilon Q' S' \mp P' \sqrt{-\epsilon
      \left\{(P')^2 + (Q')^2 + \epsilon (S')^2 \right\}}\;}{(P')^2 + (Q')^2\;} ~.
 \end{align}

To see how $E'/E$ affects the metric and the density, we write $x = E'/E$. Then
in the metric (\ref{2.1}), $g_{rr}$ is a decreasing function of $x$ provided $x
> R'/R$, while for the density (\ref{2.13}) we have
 \begin{align}
   8 \pi \rho & = \frac{6 M}{R^3} \, \frac{(M'/(3 M) - x)}{(R'/R - x)} ~,   \eqlabel{Rho2}
 \intertext{so that}
   8 \pi \pd{\rho}{x} & = - \frac{6 M}{R^3} \frac{(R'/R - M'/(3M))}{(R'/R - x)^2}
      \eqlabel{drhodx}
 \intertext{and if $x \to \pm \infty$}
   8 \pi \rho & \to \frac{6 M}{R^3} ~.   \eqlabel{rholim}
 \end{align}
Therefore at given $r$ and $t$ values, the density varies monotonically with $x
= E'/E$, but the sign of the numerator may possibly change as $R$ evolves.  If
$x$ can diverge, $\rho$ approaches a finite, positive limit.

   The metric component
 \begin{align}
   \frac{(dp^2 + dq^2)}{E^2} \eqlabel{2.19}
 \end{align}
is a 2-d surface of constant unit curvature, that is a pseudosphere%
 \footnote{
 The hyperbolic equivalent of a sphere is a right hyperboloid of revolution,
often called a pseudosphere.
 }%
 , a plane, or a sphere in Riemann or stereographic projection:
 \begin{align}
   & \epsilon = -1~,~ E > 0:~~
         \frac{(p - P)}{S} = \coth\left(\frac{\theta}{2}\right)
         \cos(\phi) ~, \nn \\
   &~~~~~~ \frac{(q - Q)}{S} = \coth\left(\frac{\theta}{2}\right) \sin(\phi) ~,
      \eqlabel{Riemprojm} \\
   & \epsilon = -1~,~ E < 0:~~
         \frac{(p - P)}{S} = \tanh\left(\frac{\theta}{2}\right)
         \cos(\phi)~, \nn \\
   &~~~~~~ \frac{(q - Q)}{S} = \tanh\left(\frac{\theta}{2}\right) \sin(\phi) ~,
      \eqlabel{Riemprojm2} \\
   & \epsilon = ~0:~~
         \frac{(p - P)}{S} = \left(\frac{2}{\theta}\right)
         \cos(\phi) ~, \nn \\
      &~~~~~~ \frac{(q - Q)}{S} = \left(\frac{2}{\theta}\right) \sin(\phi) ~,
      \eqlabel{Riemproj0} \\
   & \epsilon = +1: ~~\mbox{either}~~
         \frac{(p - P)}{S} = \cot\left(\frac{\theta}{2}\right)
         \cos(\phi) ~, \nn \\
   &~~~~~~ \frac{(q - Q)}{S} = \cot\left(\frac{\theta}{2}\right) \sin(\phi) ~.
      \eqlabel{2.42} \\
   & \mbox{or}~~~~~~~~~~~~~~~~~~~~
         \frac{(p - P)}{S} = \tan\left(\frac{\theta}{2}\right)
         \cos(\phi)~, \nn \\
      &~~~~~~ \frac{(q - Q)}{S} = \tan\left(\frac{\theta}{2}\right) \sin(\phi),
      \eqlabel{Riemprojp2}
 \end{align}

 \begin{figure}
 \parbox{80mm}{
 \includegraphics[scale = 0.55]{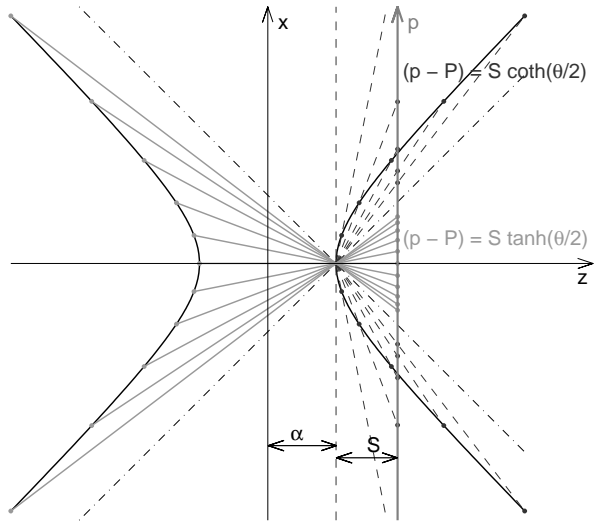}
 \caption{
 \figlabel{RieProFigH}
 \footnotesize
The Riemann projection from $(\theta, \phi)$ to $(p, q)$ coordinates for
pseudospheres ($\epsilon = -1$).  The projections of the two sheets require
different formulae, one is shown as solid grey lines, the other as dark dashed
lines.  The $45^\circ$ asymptotes that divide the projections of the two
hyperboloid sheets are shown as dot-dash lines. This and the next 2 diagrams
show only the $\phi = 0, \pi$ section, i.e. the $q = Q$ section.  For the full
projection, they should be rotated around the $z$ axis and the $q$ dimension
added.
 }
 }
 \end{figure}
 \begin{figure}
 \parbox{80mm}{
 \includegraphics[scale = 0.55]{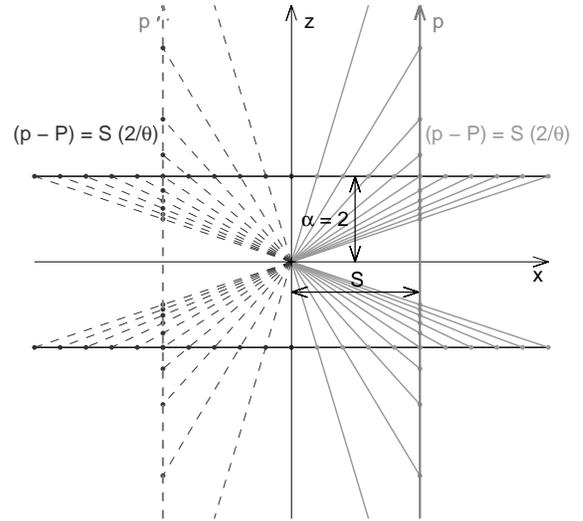} \\
 \caption{
 \figlabel{RieProFigC}
 \footnotesize
 The Riemann projection from $(\theta, \phi)$ to $(p, q)$ coordinates for semi-infinite
cylinders ($\epsilon = 0$).  Section \ref{FlatLim} gives the projection as an
inversion of the plane in a circle, which can't be illustrated as above.  The
diagram here shows the projection of a cylinder, with $\theta$ increasing along
the length of the cylinder, and each half cylinder maps to the full $(p, q)$
plane (with the same formula).
 }
 }
 \end{figure}
 \begin{figure}
 \parbox{80mm}{
 \includegraphics[scale = 0.55]{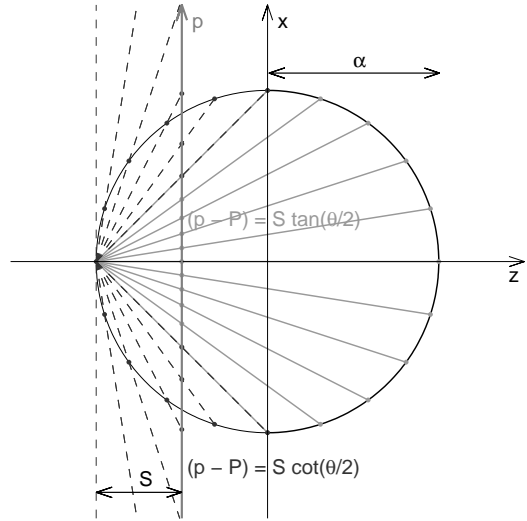}
 \caption{
 \figlabel{RieProFigS}
 \footnotesize
 The Riemann projection from $(\theta, \phi)$ to $(p, q)$ coordinates for spheres
($\epsilon = +1$).  Each of the two possible projection formulae maps the full sphere
to the plane, but only half of each is shown, one as solid grey lines, the other as
dark dashed lines.
 }
 }
 \end{figure}

 \begin{figure}
 \includegraphics[scale = 0.45]{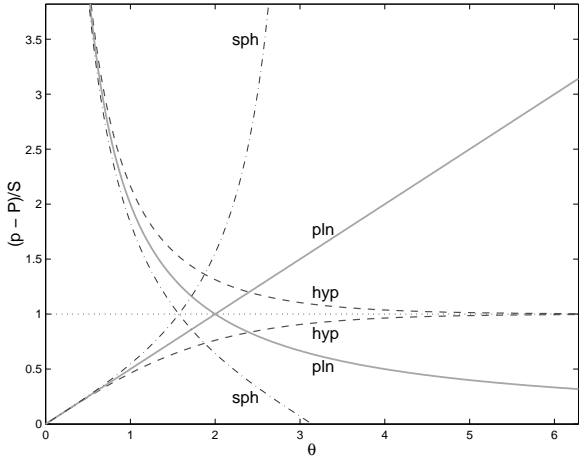}
 \caption{
 \figlabel{theta-p}
 The relation bewtween $p$ \& $\theta$ for each of the Szekeres types.  Only $(p
- P) \geq 0$ and $\theta > 0$ is shown, as rotating $\phi$ completes the
mapping.  The dark dot-dash line is for $\epsilon = +1$, the pale solid line is
for $\epsilon = 0$, and medium dashed line is for $\epsilon = -1$.
 }
 \end{figure}

   The projections are illustrated in figs \ref{RieProFigH}-\ref{RieProFigS}, and
the $\theta$-to-$p$ transformations (at $\phi = 0$) are shown in fig \ref{theta-p}.
In these diagrams, the parametric equations for spheres and right hyperboloids are
 \begin{align}
   x & = a \sin \theta \cos \phi ~,~~
         y = a \sin \theta \sin \phi ~,~~
         z = a \cos \theta ~, \nn \\
      &~~~~~~ 0 \leq \theta \leq \pi ~,~~ 0 \leq \phi \leq 2 \pi ~,
         \eqlabel{xyzs} \\
   x & = a \cos \phi ~,~~
         y = a \sin \phi ~,~~
         z = a \theta ~, \nn \\
      &~~~~~~ 0 \leq \theta \leq \infty ~,~~ 0 \leq \phi \leq 2 \pi ~,
         \eqlabel{xyzp} \\
   x & = a \sinh \theta \cos \phi ~,~~
         y = a \sinh \theta \sin \phi ~,~~
         z = a \cosh \theta ~, \nn \\
      &~~~~~~ -\infty \leq \theta \leq \infty ~,~~ 0 \leq \phi \leq 2 \pi ~,
         \eqlabel{xyzh}
 \end{align}
where the former gives the entire sphere minus one point, but the latter gives
only one sheet of the hyperboloid%
\footnote{Thus, with $\epsilon = -1$, each constant $r$ ``shell" seems to be a
hyperboloid with two ``sheets".  It will be determined later whether both these
sheets are needed or even allowed.}%
 .  Notice that, with $\theta$ \& $\phi$ ranging over the whole sphere, {\em each}
of the spherical transformations (\ref{Riemprojm}) \& (\ref{Riemprojm2}) covers
the entire $p$-$q$ plane.  (In fig \ref{RieProFigS}, only the range $0 \leq
\theta \leq \pi/2$, $\phi = 0, \pi$ has been shown for each.)  In contrast, {\em
BOTH} of the pseudospherical transformations (\ref{2.42}) \& (\ref{Riemprojp2}),
with $0 \leq \theta \leq \infty$, are required to cover the entire $p$-$q$ plane
once, each transformation mapping one of the hyperboloid sheets to the $p$-$q$
plane.  To distinguish the sheets, we choose $\theta$ to be negative on one and
positive on the other.  In the planar case, the Riemann projection can be
considered an inversion of the plane in a circle, which is hard to illustrate,
or as in fig \ref{RieProFigC} a mapping of a semi-infinite cylinder to a plane.

One might suspect that the two regions of the $(p, q)$ plane on either side of
$E = 0$ simply provide a double covering of the same surface, but this is not
the case. For the double-sheeted hyperboloid at a single $r$ value, the two
sheets are isometric to each other, the isometry transformation $(p, q) \to (p',
q')$ is
\begin{align}\label{hypinvert}
p &= P_0 + \frac {{S_0}^2 (p' - P_0)} {(p' - P_0)^2 + (q' - Q_0)^2}, \nonumber
\\
q & = Q_0 + \frac {{S_0}^2 (q' - Q_0)} {(p' - P_0)^2 + (q' - Q_0)^2},
\end{align}
where $(S_0, P_0, Q_0)$ are the values of $P$, $Q$ and $S$ in that hyperboloid.
However, for a family of hyperboloids immersed in a Szekeres spacetime, the
transformation (\ref{hypinvert}) will change the values of the functions $(P, Q,
S)$ in all other hyperboloids, and will not be an isometry. Thus, the two sheets
are distinct surfaces in spacetime.

   It is a property of the Riemann projection that for $\epsilon \geq 0$, circles
in $(p, q)$ map to circles in $(\theta, \phi)$.  Constant $\phi$ lines in
$(\theta, \phi)$ (that obviously pass through $\theta = 0$) map to straight
lines through $p = P$, $q = Q$. Circles in $(\theta, \phi)$ that pass through
$\theta = 0$, map to straight lines in $(p, q)$. See fig. \ref{planestrip} for
an example with $\epsilon = 0$.  For $\epsilon = 0$ the projection is just an
inversion of the plane in the circle of radius $\sqrt{2 S}$.

 \begin{figure}
 \begin{center}
 \includegraphics[scale=0.5]{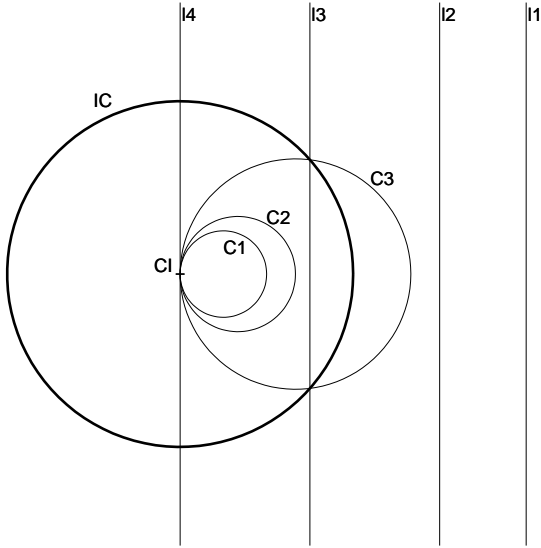}
 \caption{
 \figlabel{planestrip}
\footnotesize An inversion with respect to the circle IC centred at CI maps an
infinite straight line into a circle that passes through CI. The straight lines
$l_1$, $l_2$ and $l_3$ are mapped into the circles $C_1$, $C_2$ and $C_3$,
respectively. The straight line $l_4$ that passes through CI is mapped onto
itself, i.e. the image-circle has then an infinite radius. A strip between two
parallel straight lines is mapped into the crescent-shaped ring between their
image-circles. The ring has finite surface area except when one edge of the
strip passes through CI.
  }
 \end{center}
 \end{figure}

 Thus the factor $\epsilon$ determines whether the $p$-$q$ 2-surfaces are
pseudospherical ($\epsilon = -1$), planar ($\epsilon = 0$), or spherical
($\epsilon = +1$).  In other words, it determines the shape of the constant $r$
2-surfaces that foliate the 3-d spatial sections of constant $t$.  The function
$E$ determines how the coordinates $(p,q)$ map onto the 2-d unit pseudosphere,
plane or sphere at each value of $r$.  Each 2-surface is multiplied by factor $R
= R(t,r)$ that is different for each $r$ and evolves with time. Thus the
$r$-$p$-$q$ 3-surfaces are constructed out of a sequence of 2-dimensional
spheres, pseudospheres, or planes that are not arranged symmetrically.
Obviously, for $\epsilon \leq 0$ the area of the $(t =$ const, $r = $const$)$
surfaces could be infinite, but in the $\epsilon = +1$ case it is $4\pi R^2$.

 \section{The Effect of $\epsilon$ and $E$}
 \seclabel{RoleE}

We here analyse the role $E$ plays in these models, and contrast it with the
$\epsilon = +1$ case, in which $E'/E$ creates a dipole variation around the
constant $(t,r)$ 2-spheres.  We omit some of the detail below because very
similar calculations were done in \cite{HeKr2002}.  We assume $S > 0$.

 \subsection{Pseudospherical foliations, $\epsilon = -1$}
 \seclabel{RoleEeps-1}

Transforming (\ref{Edef}) and its derivatives using (\ref{Riemprojm}) and
(\ref{Riemprojm2}), and putting $\epsilon = -1$ , we get
 \begin{align}
   E   & = \frac{\nu \, S}{\cosh \theta - \nu} ~,
      \eqlabel{E-thetah}   \\ \nn \\
   E'  & = - \frac{S' \cosh \theta
      + \sinh \theta (P' \cos \phi + Q' \sin \phi)}{\cosh \theta - \nu} ~,
      \eqlabel{E'-thetaphih}
 \end{align}
 \begin{align}
   E'' & = - \frac{S'' \cosh \theta
         + \sinh \theta (P'' \cos \phi + Q'' \sin \phi)}{(\cosh \theta - \nu)}
         \nn \\ \nn \\
       &~~~~ + 2 \left( \frac{S'}{S} \right) \left( \frac{S' \cosh \theta
          + \sinh \theta (P' \cos \phi + Q' \sin \phi)}
          {(\cosh \theta - \nu)} \right)
         \nonumber \\ \nn \\
       &~~~~ - \frac{((S')^2 - (P')^2 - (Q')^2)}{S} ~,
         \eqlabel{E''-thetaphih}
 \end{align}
where $\nu = +1$ when $E > 0$,~ $0$ when $E = 0$~ and $-1$ when $E < 0$. The $E
= 0$ circle corresponds to $\theta \to \pm \infty$, and its neighbourhood
represents the asymptotic regions of the two sheets. It is clear that curves and
regions that intersect the $E = 0$ circle must have infinite length or area,
since
 \begin{align}
L = \int \frac{R}{E} \sqrt{dp^2 + dq^2}\; ds ~,~~~~ A = \int \int
\frac{R^2}{E^2} \, dp \, dq
 \end{align}

   The locus $E' = 0$ for all $E$ is
 \begin{align}
S' \cosh \theta + P' \sinh \theta \cos \phi + Q' \sinh \theta \sin \phi = 0 ~.
      \eqlabel{E'=0h}
 \end{align}
Writing $z = \cosh \theta$, $y = \sinh \theta \sin \phi$, $x = \sinh \theta \cos
\phi$ as the parametric locus of a unit right hyperboloid centered on $(0, 0,
0)$ in flat 3-d space, we find (\ref{E'=0h}) becomes $S' z + P' x + Q' y = 0$
which is a plane through $(0, 0, 0)$, so $E' = 0$ is the intersection of a plane
with a right hyperboloid.  In fact, (\ref{E'=0h}) is a geodesic of the $p$-$q$
2-space, as shown in appendix \ref{Geod}.

 We can write the $E' = 0$ locus as
 \begin{align}
   \tanh \theta & = \frac{- S'}{P' \cos \phi + Q' \sin \phi}
      = \frac{- d}{\cos(\phi - \phi_0)}   \eqlabel{E'=0eqh}
 \intertext{where}
   d & = \frac{S'}{\sqrt{(P')^2 + (Q')^2}\;} ~,~~~~
   \tan(\phi_0) = \frac{Q'}{P'}   \eqlabel{dphi0Def} \\
\intertext{so obviously a solution only exists if (\ref{hE'=0exists}) holds, and
only for}
   & |\cos(\phi - \phi_0)| \geq d ~.
 \end{align}

   From (\ref{E'-thetaphih}) and (\ref{E-thetah}) we find
 \begin{align}
   \frac{E'}{E} = - \nu \, \frac{S' \cosh \theta + \sinh \theta (P' \cos \phi
   + Q' \sin \phi)}{S}
   \eqlabel{ErE-thetaphi} ~,
 \end{align}
 thus $E'/E =$~constant implies $S' z + P' x + Q' y = S \times$~constant, which
is a plane parallel to the $E' = 0$ plane. The location of the extrema of $E'/E$
are found as follows
 \begin{align}
   \frac{\partial (E'/E)}{\partial \phi} & =
         \nu \, \frac{\sinh \theta (P' \sin \phi - Q' \cos \phi)}{S} = 0 \\
   & \Rightarrow~~ \tan \phi_e = \frac{Q'}{P'} ~~~~\mbox{and} \nn \\
      & \cos \phi_e = \epsilon_1 \frac{P'}{\sqrt{(P')^2 + (Q')^2}\;} ~,
         \eqlabel{TanPhiX}
 \end{align}
 \begin{align}
   \frac{\partial (E'/E)}{\partial \theta} & = 0 = \nn \\
         - \nu & \, \frac{S' \sinh \theta + \cosh \theta (P' \cos \phi + Q'
               \sin \phi)}{S} \\
   & \Rightarrow~~ \tanh \theta_e = - \frac{P' \cos \phi_e + Q' \sin
              \phi_e}{S'} \nn \\
      &~~~~~~~~ = - \epsilon_1 \frac{\sqrt{(P')^2 + (Q')^2}\;}{S'}
      ~~~~\mbox{and} \eqlabel{TanhThetaX} \nn \\
   & \cosh \theta_e = \epsilon_2 \frac{S'}{\sqrt{(S')^2 - (P')^2 - (Q')^2}\;}~.
 \end{align}
where $\epsilon_1 = \pm 1$ and $\epsilon_2 = {\rm sign}(S')$.  The extreme value
is then
 \begin{equation}
   \left( \frac{E'}{E} \right)_{\rm extreme} =
      - \epsilon_2 \,\nu \frac{\sqrt{(S')^2 - (P')^2 - (Q')^2}}{S}.
      \eqlabel{E'Eextreme}
 \end{equation}
and these extrema only exist at finite $\theta$ if
 \begin{align}
   (S')^2 > (P')^2 + (Q')^2   \eqlabel{maxE'/Eexists}
 \end{align}
 which is the opposite of (\ref{hE'=0exists}); so on a given constant $r$ shell,
either $E' = 0$ exists, or the extrema of $E'/E$ exist, but not both.  Notice
that when (\ref{maxE'/Eexists}) holds, then $E'$ does not change sign on a given
sheet, it is fixed by $\nu$ and the sign of $S'$.  It follows from
(\ref{ErE-thetaphi}) that this extremum is a maximim where $E'/E$ is negative,
and a minimum where $E'/E$ is positive.
Thus, for each constant $r$ hyperboloid, on the sheet with $E S' < 0$ (i.e. $\nu \epsilon_2
= -1$), $E'/E$ has a positive minimum and goes to $+\infty$ as $|\theta| \to \infty$,
while on the sheet with $E S' > 0$, $E'/E$ has a negative maximum and goes to $-\infty$.
The maximum and minimum are at opposite poles in the sense that
$(\theta, \phi)~\to~(-\theta, \phi + \pi)$ maps one into the other, and indeed it maps
$E'/E$ to $-E'/E$. We now specify that $\theta < 0$ on the $E < 0$ sheet (see below
(\ref{xyzh})).

{}From the foregoing considerations, if $(S')^2 > (P')^2 + (Q')^2$, then $E'/E$
is the pseudospherical equivalent of a dipole, having a negative maximum on one
sheet and a positive minimum on the other, but diverging in the asymptotic
regions of each sheet near $E = 0$.

   We see in the metric (\ref{2.1}) that $R E'/E$ is the correction to the
separation $R'$, along the $r$ curves, of neighbouring constant $r$ shells,
meaning that the hyperboloids are centered differently and are ``non
concentric".  In particular $R S' / S$ is the forward $(\theta = 0)$
displacement, and $R P' / S$ \& $R Q' / S$ are the two sideways displacements
$(\theta = \pi/2,~ \phi = 0)$ \& $(\theta = \pi/2,~ \phi = \pi/2)$.  The
shortest radial distance is where $E'/E$ is maximum.  (From a given point $(p,
q)$ on a given $r$ shell at constant $t$, the shortest distance to an
infinitesimally neighbouring $r$ shell must be along an orthogonal curve, i.e.
along constant $p$ and $q$.)

 \subsection{Planar foliations, $\epsilon = 0$}
 \seclabel{RoleEeps0}

Transforming (\ref{Edef}) and its derivatives using (\ref{Riemproj0}) and
putting $\epsilon = 0$ , we get
 \begin{align}
   E   & = \frac{2 S}{\theta^2},
      \eqlabel{E-thetap}   \\ \nn \\
   E'  & = - \frac{2(S' + \theta (P' \cos \phi + Q' \sin \phi))}{\theta^2},
      \eqlabel{E'-thetaphip}   \\ \nn \\
   E'' & = - \frac{2(S'' + \theta (P'' \cos \phi + Q'' \sin \phi))}{\theta^2}
         \nn \\ \nn \\
       & + 4 \left( \frac{S'}{S} \right) \left( \frac{S'
          + \theta (P' \cos \phi + Q' \sin \phi)}{\theta^2} \right)
         \nonumber \\ \nn \\
       & + \frac{(P')^2 - (Q')^2}{S}.
         \eqlabel{E''-thetaphip}
 \end{align}
 The $E = 0$ locus has shrunk to the point $p = P$, $q = Q$, but still
corresponds to the asymptotic regions of the plane, $\theta = \infty$.
 The locus $E' = 0$ is
 \begin{align}
   S' + P' \theta \cos \phi + Q' \theta \sin \phi = 0 ~.
      \eqlabel{E'=0p}
 \end{align}

Obviously, (\ref{E'=0p}) is a geodesic of the $p$-$q$ 2-space. We can write the
$E' = 0$ locus as
 \begin{align}
   \theta & = \frac{- S'}{P' \cos \phi + Q' \sin \phi}
      = \frac{- d}{\cos(\phi - \phi_0)}   \eqlabel{E'=0eqp}
 \end{align}
where (\ref{dphi0Def}) defines $d$ and $\phi_0$, and evidently it exists
provided
 \begin{align}
S' \neq 0 ~~~~\mbox{and}~~~~ (P' \neq 0 ~~\mbox{or}~~ Q' \neq 0) ~.
\eqlabel{pE'=0exists}
 \end{align}

   From (\ref{E'-thetaphip}) and (\ref{E-thetap}) we find
 \begin{align}
   \frac{E'}{E} = - \frac{S' + \theta (P' \cos \phi + Q' \sin \phi)}{S}
   \eqlabel{ErE-thetaphip} ~.
 \end{align}
 Thus there are no extrema of $E'/E$, and it extends to both $\pm \infty$, though
for fixed $\theta$, $\phi = \phi_0 \pm \pi$ gives the line of maximum and
minimum $E'/E$.  The behaviour found here cannot really be termed a dipole.

As before, $R E'/E$ is the correction to the ``radial" separation $R'$ of
neighbouring constant $r$ shells, and the above indicates that adjacent shells
are planes tilted relative to each other, with $\phi_0$ being the direction of
maximum tilt, but if $E'/E =$~constant they are parallel.

 \section{Regularity}

 \subsection{Pseudospherical and Planar ``Origins'', }
 \seclabel{HSPOrig}

For spherical foliations, $\epsilon = +1$, if $r = 0$ is an origin, then $R(t,0)
= 0$ for all $t$, and such origins are well understood.  The conditions on the
arbitrary functions that ensure a regular origin were given in \cite{HeKr2002}.
Specifically, the density, curvature and evolution of $R$ are all well behaved
if
 \begin{align}
   M \sim R^3 ~,&~~~~
   f \sim R^2 ~,~~~~ \nn \\
   S \sim R^n ~,~~~~
   P \sim R^n ~,&~~~~
   Q \sim R^n ~,~~~~ n \geq 0.
   \eqlabel{R=0cond}
 \end{align}
We note that the derivation of these conditions does not depend on the value of
$\epsilon$. Therefore one immediately asks whether such a locus is possible for
pseudospherical and planar foliations.

Now by (\ref{epsf>=0}) we must have $f \geq -\epsilon$ for a Lorentzian
signature, so for $\epsilon = -1$ models, $f \to 0$ is not possible.  Therefore
an ``origin" is {\em not allowed} for pseudospherical foliations.

For planar foliations, $\epsilon = 0$, $f \to 0$ is not impossible.  By
(\ref{grrFinite}) we expect
 \begin{align}
   \lim_{r \to r_O} g_{rr} = \lim_{r \to r_O}
   \frac{ \left\{ R' \left[ 1 - \frac{R E'}{R' E} \right] \right\}^2 }{f}
 \end{align}
to be finite and non-zero, and from (79) and (84) of \cite{HeKr2002} we know $R
E'/(R' E)$ is not divergent.  So, to keep $g_{rr}$ well behaved in this limit,
we require $R'/\sqrt{f}$ to be finite and non-zero, and by (\ref{R=0cond}) this
implies
 \begin{align}
   R' \sim \sqrt{f} \sim R ~~~~\Rightarrow~~~~
   R \sim e^{br} ~,~~~~ b~\mbox{constant,}
   \eqlabel{R=0condp}
 \end{align}
 while the ``radial" distance is
 \begin{align}
   s = \int \sqrt{g_{rr}}\; \, dr \sim r ~.
 \end{align}
In other words, R only asymptotically approaches zero.  Therefore there is no
real origin, but $R$, $M$ and $f$ can asymptotically approach zero. (See figure
\ref{r-theta-rRW-thRW}.)

 \subsection{Conditions for No Shell Crossings}
 \seclabel{NoShCr}

For $\rho$ to be positive, (\ref{2.13}) shows that $(M' - 3 M E'/E)$ \& $(R' - R
E'/E)$ must have the same sign.  We now consider the case where both are
positive.  Where $(M' - 3 M E'/E) \leq 0$ and $(R' - R E'/E) < 0$ we reverse the
inequalities in all the following.

 \subsubsection{Pseudospherical foliations, $\epsilon = -1$}
 \label{NoShCrPsSphFol}

   The inequality
 \begin{align}
    (M' - 3 M E'/E) \geq 0   \eqlabel{M'3E'E}
 \end{align}
must hold for all possible $p$ \& $q$, and at every $r$ value. If
(\ref{hE'=0exists}) holds so that there is an $E' = 0$ locus on each hyperboloid
sheet, then $E'/E$ varies between $\pm \infty$, diverging in the asymptotic
regions of each sheet, so the density inevitably goes negative in some regions
of every constant $r$ shell.
If however (\ref{maxE'/Eexists}) holds, so there are finite extreme values for
$E'/E$ but no loci where $E'/E = 0$, then on the sheet with $0 \leq
{(E'/E)_{\text{min}}} \leq (E'/E) < \infty$, (\ref{M'3E'E}) is violated over all
of the sheet, except near the minimum if $M'/(3M) \geq {(E'/E)_\text{min}} =
\sqrt{(S')^2 - (P')^2 - (Q')^2}\,/S$, but on the sheet with $0 \geq
{(E'/E)_\text{max}} \geq (E'/E) > -\infty$, it is always satisfied if
 \begin{align}
   \frac{M'}{3M} \geq {\left( \frac{E'}{E} \right)_\text{max}}
   =  - \frac{\sqrt{(S')^2 - (P')^2 - (Q')^2}\;}{S} ~.   \eqlabel{M'E'Emax}
 \end{align}
 It is obvious that (\ref{maxE'/Eexists}) and (\ref{M'E'Emax}) ensure (\ref{M'3E'E}),
but (\ref{M'E'Emax}) can only hold for one sheet, and on that sheet it appears
that negative $M'$ $f'$ or $R'$ are not excluded.

   Now consider the time evolution of $(R'/R - E' / E) > 0$.  Because of the above,
we only need consider the negative $E'/E$ sheet, and since $\epsilon = -1$, only
hyperbolic evolution, with $f \geq 1$, is relevant.  The argument proceeds
almost exactly as in \cite{HeKr2002}, except that the dipole term $-RE'/E$ is
everywhere positive, so it tends to relax the conditions.  Defining $\phi_4 =
\sinh\eta(\sinh\eta - \eta)/(\cosh\eta - 1)^2$ and $\phi_5 =
\sinh\eta/(\cosh\eta - 1)^2$ we have
 \begin{align}
   \frac{R'}{R} & = \frac{M'}{M}(1 - \phi_4)
      + \frac{f'}{f} \left( \frac{3}{2}\phi_4 - 1 \right)
      - \frac{f^{3/2} a'}{M} \phi_5 ~.
      \label{R'RH}
 \end{align}
Because $(1 - \phi_4)$, $(3 \phi_4 - 2)$ and $\phi_5$ are always positive, but
evolve differently with $\eta$, this argument shows that to avoid shell
crossings we require
 \begin{align}
   a' \leq 0 ~,  \eqlabel{arCondH}
 \end{align}
 and
 \begin{align}
   \frac{f'}{2f} - \frac{E'}{E} \geq 0 ~,   \eqlabel{R'/R>E'/E}
 \end{align}
 and the latter takes its strongest form at the maximum of $E'/E$, so
 \begin{align}
\frac{f'}{2f} \geq - \frac{\sqrt{(S')^2 - (P')^2 - (Q')^2}\;}{S} ~.
\eqlabel{frCondH}
 \end{align}
To confirm (\ref{maxE'/Eexists}), (\ref{M'E'Emax}), (\ref{arCondH}) and
(\ref{frCondH}) are sufficient, we use equations (99) and (100) of
\cite{HeKr2002}, write $X = \left|{(E'/E)_\text{max}}\right|$ so
(\ref{M'E'Emax}) and (\ref{frCondH}) become $M'/(3M) = - X + \alpha$ and
$f'/(2f) = -X + \beta$ with $\alpha$ and $\beta$ non-negative,
and thus obtain (\ref{R'RH}) again and
 \begin{align} \label{R'RH1}
   \frac{R'}{R} & = - X + 3 \alpha (1 - \phi_4) + \beta (3 \phi_4 - 2)
      - \frac{f^{3/2} a'}{M} \phi_5 ~, \\
   & \geq -X ~,
 \end{align}
 as required.  This also means
 \begin{align}
\frac{R'}{R} - \frac{M'}{3M} & \geq  (\beta - \alpha) (3 \phi_4 - 2) -
\frac{f^{3/2} a'}{M} \phi_5 ~.
      \eqlabel{R'RM'3Mev}
\intertext{so the numerator of (\ref{drhodx}) is negative for all $\eta$ if
there are no shell crossings and}
   & \frac{f'}{2 f} \geq \frac{M'}{3 M} ~,   \eqlabel{R'RM'3Mpos}
 \end{align}
otherwise it can change sign from negative to positive as $\eta$ increases or if
$a' = 0$ it goes from zero to negative.

Thus we see that only one of the hyperboloid sheets can be free of shell crossings,
and it must have a minimum in $(R'- R E'/E)$ and $(M'- M E'/E)$.

 \subsubsection{Planar foliations, $\epsilon = 0$}

   By (\ref{ErE-thetaphip}) and the discussion following (\ref{E'=0p}) we have
 \begin{align}
   \frac{R'}{R} - \frac{E'}{E} & =
   \frac{R'}{R} + \frac{S' + \theta (P' \cos \phi + Q' \sin \phi)}{S} ~,
 \end{align}
so adjacent shells are like tilted planes and inevitably they must intersect on
the straight line
 \begin{align}
   \theta & = \frac{-(S' + S R' / R)}{P' \cos \phi + Q' \sin \phi}
 \end{align}
 creating shell crossings, except when
 \begin{align}
   P' = 0 ~,~~ Q' = 0 ~,   \eqlabel{P'Q'Condp}
 \end{align}
 and
 \begin{align}
   \frac{R'}{R} \geq \frac{-S'}{S} ~.   \eqlabel{R'S'Condp}
 \end{align}
Condition (\ref{P'Q'Condp}) ensures the shells are parallel, while
(\ref{R'S'Condp}) can be converted to
 \begin{align}
   \frac{R'}{R} \geq 0   \eqlabel{R'S'Condp2}
 \end{align}
because $S$ can be absorbed into other functions, as shown in equation
(\ref{2.12}). Effectively then we require
 \begin{align}
   S' = P' = Q' = E' = 0 ~.   \eqlabel{S'P'Q'E'Condp}
 \end{align}
 and the remaining conditions follow exactly as in \cite{HeKr2002} or
\cite{HelLak85}.

  The no shell crossing conditions for $\epsilon \leq 0$ are summarised in Table 1.
It is a continuation of Table 1 in Sec. VI of Ref. \cite{HeKr2002}, which summarised
those conditions for $\epsilon = +1$.

 \subsection{Regular Maxima and Minima}
 \seclabel{RegMaxMin}

   We already know that spherical foliations can have regular extrema $r = r_m$,
where $R'(t, r_m) = 0$, and we consider this possibility for other $\epsilon$
values.  The case of both $(M' - 3 M E'/E)$ \& $(R' - R E'/E)$ being zero may
occur momentarily at isolated locations as $R$ evolves, but for a given $r =
r_m$, the no shell crossing considerations give
 \begin{align}
   R' = M' = f' = a' = S' = P' = Q' = 0 ~,   \eqlabel{R'=0Cond}
 \end{align}
 since they must hold at all times, and for all $p$ \& $q$.  In order for the
metric and the density to have well behaved limits as $r_m$ is approached, we
require
 \begin{align}
   \sqrt{g_{rr}} = \frac{R' - R E'/E}{\sqrt{\epsilon + f}\;}
      & ~\to~ L ~,~~~~ 0 < L < \infty   \eqlabel{R'0grrlim} \\
   4 \pi \rho R^2 = \frac{M' - 3 M E'/E}{R' - R E'/E}
      & ~\to~ N ~,~~~~ 0 \leq N < \infty   \eqlabel{R'0rholim}
 \end{align}
 and we obtain all the results of section VI of \cite{HeKr2002}, which was done
for general $\epsilon$.
As noted there, we must replace $M'$ with $\lim_{r \to
r_m} M'/\sqrt{\epsilon + f}\,$, and similarly for all 6 arbitrary functions, in
all the no shell crossing conditions;
and to ensure these limits exist as well as avoid a surface layer at $r_m$ we
require
 \begin{align}
   f = -\epsilon ~.   \eqlabel{fNoSfLayer}
 \end{align}
With pseudospherical foliations this just means $f = +1$ at an extremum, but
with planar foliations, we already saw in section \ref{HSPOrig} that $f = 0$ is
not possible, and can only be approached asymptotically,
so spatial extrema of $R$ cannot occur when $\epsilon = 0$.

 \begin{widetext}
 \begin{center}
{\footnotesize Table 1.~~ Summary of the conditions for no shell crossings or
surface layers.}
 \setlength{\tabcolsep}{2mm}
 \renewcommand{\arraystretch}{1}
 \begin{tabular}{|c|c|c|c|c|}
 \hline \hline
 $\epsilon$ & $R'$ & $f$ & $S'$ & $M'$~,~~$f'$~,~~$a'$~,~~$P'$~,~~$Q'$ \\
 \hline \hline
   $= -1$ & $> 0$ & $\geq 1$ & $E S' >0$
      & ~\parbox{50mm}{${}$ \\[1mm]
                      $(S')^2 > (P')^2 + (Q')^2$ \\[1mm]
                      $\frac{M'}{3 M} \geq - \frac{\sqrt{(S')^2 - (P')^2 -
                                      (Q')^2}\;}{S}$ \\[1mm]
                      $\frac{f'}{2 f} \geq - \frac{\sqrt{(S')^2 - (P')^2 -
                                      (Q')^2}\;}{S}$ \\[1mm]
                      $a' \leq 0$} \\[13mm]
 \cline{2-5}
   & $= 0$ & $= 1$    & $S' = 0$
      & \parbox{50mm}{${}$ \\[1mm]
                      $M' = 0$~,~~ $f' = 0$~,~~ $a' = 0$~, \\
                      $P' = 0$~,~~ $Q' = 0$} \\[5mm]
 \cline{2-5}
   & $< 0$ & $\geq 1$ & $E S' < 0$
      & \parbox{50mm}{${}$ \\[1mm]
                      $(S')^2 > (P')^2 + (Q')^2$ \\[1mm]
$\frac{M'}{3 M} \leq + \frac{\sqrt{(S')^2 - (P')^2 - (Q')^2}\;}{S}$ \\[1mm]
$\frac{f'}{2 f} \leq + \frac{\sqrt{(S')^2 - (P')^2 - (Q')^2}\;}{S}$ \\[1mm]
                      $a' \geq 0$} \\[13mm]
 \hline \hline
   $= 0$ & $> 0$ & $\geq 0$ & $= 0$
      & ~\parbox{50mm}{${}$ \\[1mm]
                      $M' \geq 0$~,~~ $f' \geq 0$~,~~ $a' \leq 0$~, \\[1mm]
                      $P' = 0$~,~~ $Q' = 0$} \\[5mm]
 \cline{2-5}
   & $= 0$ & $= 0$    & $= 0$
      & \parbox{50mm}{${}$ \\[1mm]
                      $M' = 0$~,~~ $f' = 0$~,~~ $a' = 0$~, \\
                      $P' = 0$~,~~ $Q' = 0$} \\[5mm]
 \cline{2-5}
   & $< 0$ & $\geq 0$ & $= 0$
      & \parbox{50mm}{${}$ \\[1mm]
                      $M' \leq 0$~,~~ $f' \leq 0$~,~~ $a' \geq 0$~, \\[1mm]
                      $P' = 0$~,~~ $Q' = 0$} \\[5mm]
 \hline
 \end{tabular}
 \end{center}
 \end{widetext}

 \subsection{Density: Extrema, Asymptotics, and Evolution}
 \seclabel{AsymptRho}

Considering the density (\ref{2.13}) with hyperbolic and parabolic evolution
(\ref{hypevRS})-(\ref{evoRflat}) and $R'/R$ given by (\ref{R'RH1}) and section
V.B.1 of \cite{HeKr2002}, we have
 \begin{align}
   a' \neq 0:~~ \frac{R'}{R} & \to \infty ~,~~
      & \rho & \to \frac{M' - 3 M E'/E}{4 \pi R^2 R'} \\
   a' = 0:~~ \frac{R'}{R} & \to \frac{M'}{3M} ~,~~
      & \rho & \to \frac{3 M}{4 \pi R^3} = \rho_\text{LT~early}
 \end{align}
 at early times, $\eta \to 0$, while at late times, $\eta \to \infty$,
 \begin{align}
   \frac{R'}{R} & \to \frac{f'}{2f} ~,~~~~
      & \rho & \to \frac{M' - 3 M E'/E}{4 \pi R^3 (f'/(2 f) - E'/E)} ~.
 \end{align}
 Therefore, the effect of $E'/E$ only disappears near a simultaneous bang.

We saw in section \ref{RoleEeps-1} that when $\epsilon = -1$, $E'/E$ acts like the
pseudospherical equivalent of a dipole, and section \ref{NoShCr} showed $E'/E$ must be
negative but rise to a maximum somewhere.  By (\ref{drhodx}) and (\ref{R'RM'3Mev}) the
density $\rho$ decreases monotonically with $E'/E$ if (\ref{R'RM'3Mpos}) holds,
otherwise it can change to a monotonic increase as time passes.  So we conclude that,
on shells where (\ref{R'RM'3Mpos}) holds, the density is minimum where $E'/E$ is
maximum and the shell separation minimum, and vice-versa%
\footnote{It is amusing to note that the $M' = 0$, $M > 0$, case could be called
a ``bare dipole".  Of course, this case suffers from shell crossings and
negative densities.}%
.  On shells where it doesn't hold, the initial density minimum can evolve into
a maximum.  This holds for hyperbolic evolution with any $\epsilon$ value.

 \section{The Case of $\epsilon = 0$}

   Ironically, the $\epsilon = 0$ case is the most tricky to understand.
Below we
consider the quasi-planar case in two ways; as a complete manifold with planar
foliation, and as a boundary surface between a region having a spherical
foliation and one having a pseudospherical foliation.

 \subsection{The Quasi-Planar Manifold}
 \seclabel{QPlanarManifold}

   It is difficult to interpret the Szekeres spacetime in which all the $(p, q)$
subspaces are flat, even in the limit $E' = 0$, when the spacetime becomes plane
symmetric. As seen from section \ref{3spaces}, the value $f = 0$ is not
admissible, as it makes both the metric and the curvature singular. Thus, the
quasi-planar case does not admit flat 3-dimensional subspaces, so this
case cannot provide a foliation of 3-d Euclidean space, such as the construction
of section \ref{flatmodel}.

 \subsection{The Quasi-Planar Szekeres Metric as  a Limit}
 \seclabel{PlanLim}

We here show that the planar metric can be viewed as the limit of the other two
at large $R$.  We pay particular attention to the limit of the spherical case,
with which we are more familiar.  In spherical coordinates, if $R$ is large (at
a fixed, finite $t$), the region near $\theta = 0$ looks like cylindrical
coordinates, and in the limit as $R$ diverges, the constant $r$ surfaces are
effectively planar.  In this limit pseudospherical
coordinates also look cylindrical.  We need to find a transformation that will
allow this limit but keep all physical quantities well behaved.  Let $\omega$ be
a large quantity that goes to $\infty$ in the limit, then the transformation
 \begin{align}
   & M \to \omega^3 \o{M} ~,~~~~
      f \to \omega^2 \o{f} ~,~~~~
      a \to \o{a} ~,~~~~ \nn \\
   & S \to \omega^{-1} \o{S} ~,~~~~
      P \to \o{P} ~,~~~~
      Q \to \o{Q} ~, \nn \\
   & t \to \o{t} ~,~~~~
      r \to \o{r} ~,~~~~
      p \to \o{p} ~,~~~~
      q \to \o{q} ~,~~~~ \nn \\
   & \theta \to \omega^{-1} \o{\theta} ~,~~~~
      \phi \to \o{\phi} ~, \nn \\
   & \eta \to \o{\eta} ~,~~~~
      R \to \omega \o{R} ~,~~~~
      E \to \omega \o{E} ~,~~~~ \nn \\
   & \rho \to \o{\rho}
      ~,
      \eqlabel{omegaLim}
 \end{align}
 results in
 \begin{align}
   E = & \frac{S}{2} \left\{ \left( \frac{p - P}{S} \right) ^2
         + \left( \frac{q - Q}{S} \right) ^2 + \epsilon \right\} \nn
 \end{align}
 \begin{align}
   \to &~~
      \o{E} = \frac{\o{S}}{2} \left\{ \left( \frac{\o{p} - \o{P}}{\o{S}} \right) ^2
         + \left( \frac{\o{q} - \o{Q}}{\o{S}} \right) ^2\right\} ~, \\
   \frac{(p - P)}{S} & = \cot \left( \frac{\theta}{2} \right) \cos(\phi)
      \to
      \frac{(\o{p} - \o{P})}{\o{S}} = \frac{2}{\o{\theta}} \cos(\o{\phi}) ~, \\
   \frac{(q - Q)}{S} & = \cot \left( \frac{\theta}{2} \right) \sin(\phi)
      \to
      \frac{(\o{q} - \o{Q})}{\o{S}} = \frac{2}{\o{\theta}} \sin(\o{\phi}) ~, \\
   \frac{E'}{E} & = \frac{- \{ S' \cos\theta + \sin\theta (P' \cos\phi + Q' \sin\phi) \}}{S} \nn \\
      & \to \frac{E'}{E} = \frac{- \{ S' + \theta (P' \cos\phi + Q' \sin\phi) \}}{S}
 \end{align}
 \begin{align}
   \frac{\left( R' - \frac{R E'}{E} \right)^2 \, dr^2}{\epsilon + f} & \to
      \frac{\left( \o{R}' - \frac{\o{R} \o{E}'}{\o{E}} \right)^2 \, d\o{r}^2}{\o{f}} ~,
      \eqlabel{grrPLim} \\
   R^2 \, \sin^2 \theta \, d\phi^2 & \to
      \o{R}^2 \, \o{\theta}^2 \, d\o{\phi}^2 ~,
 \end{align}
 \begin{align}
   8 \pi \rho & = \frac{2 \left( M' - 3 M E' / E \right)}{R^2 \left( R' - R E' / E \right)} \nn \\
      & \to 8 \pi \o{\rho} = \frac{2 \left( \o{M}' - 3 \o{M} \o{E}' / \o{E} \right)}
         {\o{R}^2 \left( \o{R}' - \o{R} \o{E}' / \o{E} \right)} ~,
 \end{align}
 \begin{align}
   \dot{R}^2 = \frac{2 M}{R} + f & \to
      \dot{\o{R}}^2 = \frac{2 \o{M}}{\o{R}} + \o{f} ~, \\
   R = \frac{M}{f} (\cosh \eta - 1) & \to
      \o{R} = \frac{\o{M}}{\o{f}} (\cosh \o{\eta} - 1) ~, \\
   t - a = \frac{M}{f^{3/2}} (\sinh \eta - \eta) & \to
      \o{t} - \o{a} = \frac{\o{M}}{\o{f}^{3/2}}
      (\sinh \o{\eta} - \o{\eta}) ~.
      \eqlabel{taEqPlLim}
 \end{align}
 Thus we have exactly the planar Szekeres metric, with all the correct matter
content and dynamics.  Another way of looking at this transformation is that we
have effectively taken an infinitesimal region near $\theta = 0$ at finite $r$
and blown it up to finite size.
Note that $f$ must diverge, so while an elliptic model can have infinite $R$
\cite{HelLak85}, it cannot have this limit.

 \subsection{The flat limit of the Riemann projection}
 \seclabel{FlatLim}

Equations (\ref{Riemproj0}) show that the transformation from the $(p, q)$
coordinates to the $(\theta, \phi)$ coordinates involves an inversion of the
coordinate plane in the circle of radius $\sqrt{2S}$: a point at a distance $u =
\sqrt{(p - P)^2 + (q - Q)^2}\,$ from $(P, Q)$ is mapped into a point at a
distance $\theta = \sqrt{2S}/u$, so that the product $\theta u$ is the same for
all point-image pairs. The problem is thus to set up the two other mappings in
such a way that in the limit of zero curvature (infinite radius) of the sphere
or hyperboloid they go over into an inversion of the plane%
 \footnote{
 In terms of the limit of the preceding subsection, we have $S \to \o{S}/\omega$
and $\theta \to \o{\theta}/\omega$ as $\omega \to \infty$, which does ensure
(\ref{2.42}) and (\ref{Riemprojm}) go to (\ref{Riemproj0}).}%
 .

The characteristic property of the inversion is that the inversion circle
remains invariant. The first question is thus: is it possible that in the
other two projections a circle in the curved surface is mapped into a circle
of the same radius in the plane? This must be answered separately for the
sphere and the hyperboloid, and we now proceed to this consideration.

\subsubsection{The quasi-spherical model}

For the quasi-spherical model, the projection is illustrated in Fig.
\ref{sphpreproj}. The radius of the sphere is $\alpha$, the point on the sphere
that is being mapped has the polar coordinate $\theta$ and the projection plane
$PP$ is at the distance $S$ from the projection pole $O$. The image-point in the
plane is at the distance $p$ from the axis. (We set $P$ and $Q$ to zero, since
their values are unimportant for any one shell.)

 \begin{figure}
 \parbox{80mm} {
 \hspace*{-7mm}
 \includegraphics[scale = 0.7]{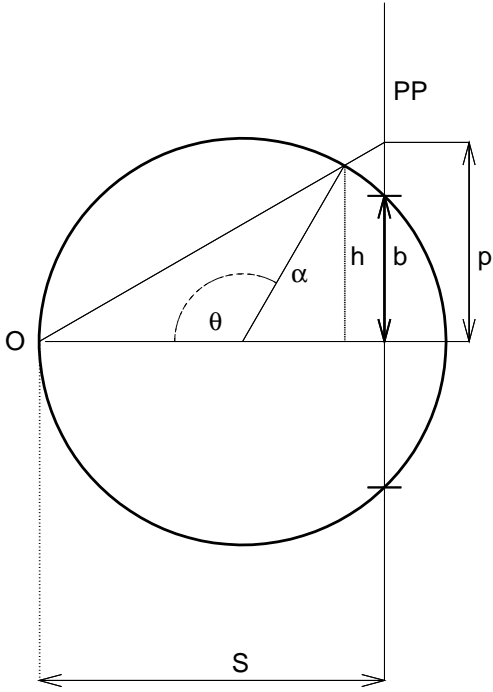}
 \caption{
 \figlabel{sphpreproj}
 \footnotesize
The Riemann projection of a sphere on a plane.  If the plane intersects the
sphere, the circle of intersection has radius $b = \sqrt{\alpha^2 - (S -
\alpha)^2}$ and it is invariant since it is mapped onto itself in the
projection. For the projection shown here, however, in the limit $\alpha \to
\infty$ with $b$ held constant, an identity mapping results, not an inversion. }
 }
 \end{figure}

If $PP$ intersects the sphere so that the radius of the intersection circle is
$b$, then points on the sphere left of $PP$ map to points on the plane outside
that circle, and vice-versa. This will become the invariant circle of the
inversion in the limit.  For a given $b$ value, there are two possible locations
for the plane, $S = \alpha \pm \sqrt{\alpha^2 - b^2}$. The ``$+$" configuration,
shown in Fig. \ref{sphpreproj} is, however, unsuitable for the limit of infinite
radius, because the part of the sphere right of $PP$ is mapped onto the inside
of the circle in the plane, and in the limit $\alpha \to \infty$ we will not get
an inversion, but an identity mapping.

Therefore we proceed to the ``$-$" configuration,
\begin{equation}\label{6.12}
S = \alpha - \sqrt{\alpha^2 - b^2},
\end{equation}
shown in Fig. \ref{spherproj}. We begin with a sphere of radius $\alpha$, and
increase $\alpha$ while moving the center of the sphere to the right in such a
way that all spheres intersect the plane $PP$ along the same circle of radius
$b$. We have $\tan (\theta/2) = S/p$, and so
\begin{equation}\label{6.13}
h = \frac {2\alpha Sp} {p^2 + S^2} = \frac {2\alpha p \left(\alpha -
\sqrt{\alpha^2 - b^2}\right)} {\left(\alpha - \sqrt{\alpha^2 - b^2}\right)^2 +
p^2}.
\end{equation}
We now apply the identity $\alpha - \sqrt{\alpha^2 - b^2} = b^2 / \left(\alpha +
\sqrt{\alpha^2 - b^2}\right)$ and obtain
\begin{equation}\label{6.14}
h = \frac {2\alpha b^2p} {\left(\alpha + \sqrt{\alpha^2 - b^2}\right)
\left[\left(\frac {b^2} {\alpha + \sqrt{\alpha^2 - b^2}}\right)^2 + p^2\right]}
\llim{\alpha \to \infty} \frac {b^2} p,
\end{equation}
which is indeed an inversion in the circle of radius $b$. Fig. \ref{spherproj}
shows also the trajectory of the projected point on the sphere as $\alpha \to
\infty$, while $p$ is kept fixed. That trajectory is a circle of radius $(1/2)
(b^2/p - p)$ with the centre at $(z, x) = (0, (p + b^2/p)/2)$. As should be
expected, the circle degenerates to a point when $p = b$, and its radius becomes
infinite when $p \to 0$.

 \begin{figure}
 \begin{center}
 \includegraphics[scale=0.5]{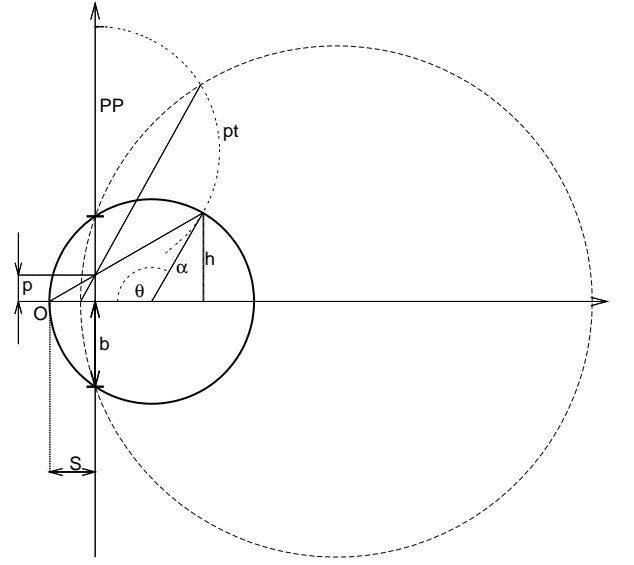}
 \caption{
 \figlabel{spherproj}
 \footnotesize
The Riemann projection of a sphere that goes over into the inversion of the
plane in the limit $\alpha \to \infty$. The meaning of the symbols is the same as
in Fig. \ref{sphpreproj}. The limit is taken in such a way that the circle of
intersection, of radius $b$, remains the same as the radius of the sphere $\alpha$
goes to infinity. One of the larger spheres is shown. If the point in the
plane at the distance $p$ from the axis is kept constant for all spheres, then
its image on the various spheres will follow the circle arch $pt$, and in the
limit $\alpha \to \infty$ the image will land in the plane, at the distance $b^2/p$
from the axis -- i.e. the limiting plane undergoes an inversion. This result is
derived in the text. }
 \end{center}
 \end{figure}

The $S$ of the flat case is actually the $b = \sqrt{S (2\alpha - S)}$ of the
spherical case, and $b$ and $h$ correspond to $\o{S}$ and $\o{\theta}$ in
(\ref{omegaLim}).

\subsubsection{The Quasi-Pseudospherical Model}

As with the spherical case, the hyperboloid and the plane must intersect for an
invariant circle to exist, and if the pole of projection is not placed in the
same sheet of the hyperboloid as the invariant circle, then in the limit of zero
curvature an identity instead of an inversion results. The case that gives the
inversion, is shown in Fig. \ref{hyperproj}.

 \begin{figure}
 \begin{center}
 \includegraphics[scale=0.5]{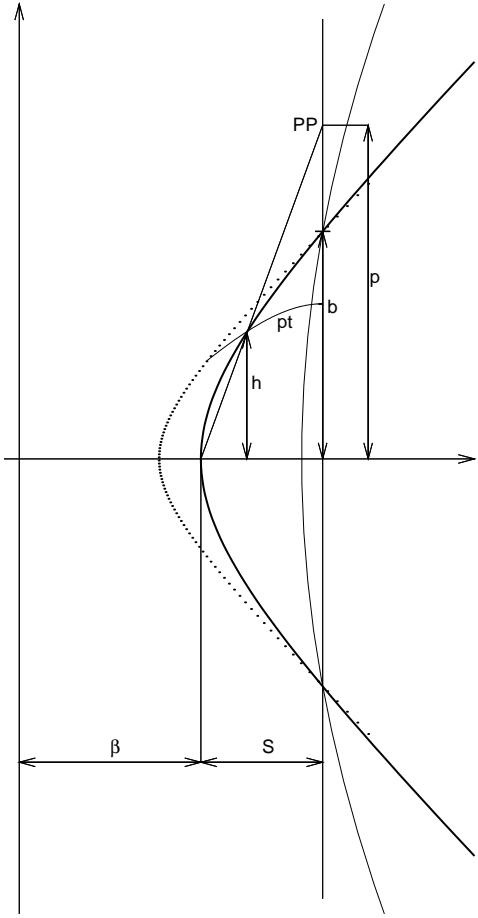}
 \caption{
 \figlabel{hyperproj}
 \footnotesize
The Riemann projection of a hyperboloid that goes over into the inversion of the
plane $PP$ in the limit $\alpha \to \infty$. Three hyperbolae (intersections of
hyperboloids with the plane of the figure) are shown. The parameter $\alpha$ has
the smallest value on the hyperbola with the leftmost vertex and largest for the
hyperbola with the rightmost vertex. The calculations in the text are done for
the middle hyperbola. As $\alpha$ increases, and $\beta = \alpha - S$ increases,
the hyperbola is shifted right so that the circle of intersection of the
hyperboloid with the plane $PP$ is always the same and has radius $b$. The curve
$pt$ is the trajectory followed by a point on the hyperboloids as $\alpha \to
\infty$, while the image point in the plane $PP$ is kept at the same distance
$p$ from the axis. All hyperbolae are right (their asymptotes are inclined at
$45^{\circ}$ to the $x$-axis); those that look wider open are simply magnified.
  }
 \end{center}
 \end{figure}

We will increase $\alpha \to \infty$, but will shift the hyperboloids
so that they intersect the plane of projection $PP$ always along the same
circle of radius $b$. Therefore we must have
\begin{equation}\label{6.15}
S = \sqrt{\alpha^2 + b^2} - \alpha ~.
\end{equation}
(We do not consider $S = \sqrt{\alpha^2 + b^2} + \alpha$ which does not lead to
inversion.) As expected, in the limit $\alpha \to \infty$ we get $S \to 0$.
Using this we get
\begin{equation}\label{6.16}
p = h\ \frac {\sqrt{\alpha^2 + b^2} - \alpha} {\sqrt{\alpha^2 + {h}^2} - \alpha}
\llim{\alpha \to \infty} \frac {b^2} h ~,
\end{equation}
which is an inversion in the circle of radius $b$.

\subsection{Joining Spherical and Pseudospherical Foliations at a Planar
Boundary}
 \seclabel{JoinSphHyp}

Suppose in an $\epsilon = +1$ Szekeres metric we let the radius of the constant
$r$ spheres diverge, so they become effectively planar at some $r$ value, and
similarly in an $\epsilon = -1$ Szekeres metric we let the ``radius" of the
hyperboloids diverge at some $r$ value. Then the the two metrics can be joined
at their planar boundaries, provided we carry out the planar limits of
(\ref{omegaLim})-(\ref{taEqPlLim}), as is easily verified by calculating the
junction conditions.

 \section{Comparison with Allied Metrics}

 \subsection{Foliations of the Robertson-Walker Metric}
 \seclabel{HSRW}

To better understand the $\epsilon = -1$ and $\epsilon = 0$ Szekeres foliations,
we first look at the simplest possible cases --- the homogeneous ones; i.e the
Robertson-Walker (RW) metric with planar and pseudospherical
foliations.  The RW metric in standard coordinates is
 \begin{align}
   ds^2 = - dt^2 + {\cal S}^2(t) \Bigg\{ \frac{d\tilde{r}^2}{1 - k \tilde{r}^2}
   + \tilde{r}^2 (d\vartheta^2 + \sin^2 \vartheta \, d\phi^2) \Bigg\} ~,
 \end{align}
and the spherical foliations ($\epsilon = +1$) obtained with $\tilde{r} = \sin(r)$,
$\tilde{r} = r$ and $\tilde{r} = \sinh(r)$ for $k = +1$, $k = 0$ and $k = -1$
respectively are familiar.  For $k = -1$ in particular, we have
 \begin{align}
   \tilde{r} & = \sinh(r_S) ~,~~~~ \vartheta = \theta_S \\
   \to~~ ds^2 & = - dt^2 + {\cal S}^2(t) \big\{ dr_S^2 \nn \\
      &~~~~ + \sinh^2(r_S) (d\theta_S^2 + \sin^2(\theta_S) \, d\phi^2) \big\} ~,
         \eqlabel{RWmetric-k-1eps+1} \\
   \to~~ R_S & = {\cal S} \sinh(r_S) ~,~~ f_S = + \sinh^2(r_S) ~,~~ \nn \\
      & M_S = M_0 \sinh^3(r_S) ~.   \eqlabel{RWRfMk-1eps+1}
 \end{align}
 For planar foliations, $\epsilon = 0$, with $k  = 0$ we obtain
 \begin{align}
   \tilde{r} & = \sqrt{r^2 + \theta^2}\; ~,~~~~ \vartheta = \tan^{-1}
   \frac{\theta}{r} \\
   \to~~ ds^2 & = - dt^2 + {\cal S}^2(t) \big\{ dr_{P0}^2
      + (d\theta_{P0}^2 + \theta_{P0}^2 \, d\phi^2) \big\} ~,
      \eqlabel{RWmetric-k0eps0} \\
   \to~~ R_{P0} & = {\cal S} ~,~~ f_{P0} = 0 ~,~~ M_{P0} = M_0 ~;
      \eqlabel{RWRfMk0eps0}
 \end{align}
 while with $k = -1$ we get
 \begin{align}
   r_P & = \ln(\cosh r_S + e_2 \sinh r_S \cos \theta_S) ~,~~ e_2 = \pm 1
      \eqlabel{rPStransf} \\
   \theta_P & = \frac{e_1 \sinh r_S \sin \theta_S}{(\cosh r_S
      + e_2 \sinh r_S \cos \theta_S)} ~,~~ e_1 = \pm 1
      \eqlabel{thetaPStransf} \\
   \sinh r_S & = \sqrt{ \frac{1}{4} \left( e^{r_P} (\theta_P^2 + 1)
      + e^{-r_P} \right)^2 - 1}\; \nn \\
   & = \sqrt{ \left( \frac{1}{2} \left( e^{r_P} (\theta_P^2 - 1)
      + e^{-r_P} \right) \right)^2 + \left( \theta_P e^{r_P} \right)^2}\;
      \eqlabel{rSPtransf}
 \end{align}
 \begin{align}
   e_1 \sin \theta_S & = \frac{\theta_P e^{r_P}}
      {\sqrt{ \left( \frac{1}{2} \left( e^{r_P} (\theta_P^2 - 1)
      + e^{-r_P} \right) \right)^2 + \left( \theta_P e^{r_P} \right)^2}\;}
      \eqlabel{sinthetaSPtransf} \\
   e_2 \cos \theta_S & = \frac{- \left( \frac{1}{2} \left( e^{r_P}
      (\theta_P^2 - 1) + e^{-r_P} \right) \right)} {\sqrt{ \left(
      \frac{1}{2} \left( e^{r_P}(\theta_P^2 - 1) + e^{-r_P} \right) \right)^2
      + \left( \theta_P e^{r_P} \right)^2}\;}
      \eqlabel{costhetaSPtransf}
 \end{align}
 \begin{align}
   \to~~ ds^2 & = - dt^2 + {\cal S}^2(t) \big\{ dr_P^2
      + e^{2 r_P} (d\theta_P^2 + \theta_P^2 \, d\phi^2) \big\} ~,
      \eqlabel{RWmetric-k-1eps0} \\
   \to~~ R_P & = {\cal S} e^{r_P} ~,~~ f_P = + e^{2 r_P}
      ~,~~ M_P = M_0 e^{3 r_P} ~,   \eqlabel{RWRfMk-1eps0}
 \end{align}
 For pseudospherical foliations, $\epsilon = -1$ \& $k = -1$, we find
 \begin{align}
   \tilde{r} & = \sqrt{\sinh^2(r_H) + \cosh^2(r_H) \sinh^2(\theta_H)} ~, \nn \\
   \vartheta & = \tan^{-1} \left( \frac{\sinh(\theta_H)}{\tanh(r_H)} \right) \\
   \to~~ ds^2 & = - dt^2 + {\cal S}^2(t) \big\{ dr_H^2 \nn \\
 &~~~~ + \cosh^2(r_H) (d\theta_H^2 + \sinh^2(\theta_H) \, d\phi^2) \big\} ~,
         \eqlabel{RWmetric-k-1eps-1} \\
   \to~~ R_H & = {\cal S} \cosh(r_H) ~,~~ f_H = + \cosh^2(r_H) ~,~~ \nn \\
         & M_H = M_0 \cosh^3(r_H) ~.   \eqlabel{RWRfMk-1eps-1}
 \end{align}
 Using the Riemann transformations, each of these can be converted to Szekeres
form, but we are here interested in understanding the relationship between
different foliations.  The relationship between $(r_H, \theta_H)$, $(r_P,
\theta_P)$  and $(r_S, \theta_S)$ for the $k = -1$ foliations is illustrated in
fig \ref{r-theta-rRW-thRW}, which plots $\tilde{r}$ and $\vartheta$
as polar coordinates on the plane.  (This compression of a negatively curved
2-surface onto the plane naturally creates some distortion.)

   We now consider time sections $t = t_0$ of the $k = -1$ RW model.  Each constant
$r_S$ 2-surface has the geometry of a sphere.  The function $\sin r_S$ has a
zero (at $r_S = 0$), a maximum (at $r_S = \pi/2$), and another zero (at $r_S =
\pi$), which are features of a closed surface in spherical coordinates.

   Each constant $r_P$ 2-surface has the geometry of a plane, but in order to embed
the plane into a negatively curved 3-space, it has to bend round so that the
circumference $e^{r_P} \theta_P \Delta \phi$ does not increase too fast compared
with the radius $e^{r_P} \theta_P$.  The constant $\theta_P$ surfaces are horns
that flare out rapidly, and even bend backwards to stay orthogonal to the
constant $r_P$ planes.  The coordinates on each plane are magnified by the
factor $e^{r_P}$ that is nowhere zero, suggesting that a plane foliation of a
negatively curved space has $R(t_0, r_P)$ decaying asypmtotically towards zero
in one direction.

   Each constant $r_H$ 2-surface has the geometry of one sheet of a two-sheeted
right hyperboloid of revolution.  The function $\cosh r_H$ has a minimum (at
$r_H = 0$), which is suggestive that a ``natural" way to cover such a negatively
curved manifold with hyperboloids is to have $R(t_0, r_H) > 0$, but going
through a minimum.

   In the spherical foliation, $M$ and $f$ are constant on spheres, are zero at an
origin, and reach a maximum where $R' = 0$. In the pseudospherical foliation,
the corresponding $M$ and $f$ functions are constant on
completely different surfaces, have a minimum where $R' = 0$, and are nowhere
zero. In the planar foliation, $M$ and $f$ are again constant on different
surfaces, have no origin and no extremum, but asymptotically approach zero.
Despite the apparently very different descriptions, these 3 foliations of the $k
= -1$ case describe exactly the same metric with the same behaviour.  In each
case they obey
 \begin{align}
   \dot{\cal S}^2 = \frac{2 M_0}{{\cal S}^2} - k ~,~~
   4 \pi \rho = \frac{3 M_0}{S^3} ~,~~ M_0 = \frac{4 \pi S_0^3 \rho_0}{3} ~,
   \eqlabel{SdotSq}
 \end{align}
 where $\rho_0$ and $S_0$ are constants.
(Due to the homogeneity, the features of $R$, $M$ \& $f$ such as the origin in
(\ref{RWRfMk-1eps+1}) or the minimum in (\ref{RWRfMk-1eps-1}) are not special
locations, as a transformation could move them to any position.)

 \begin{figure}
 \includegraphics[scale = 0.6]{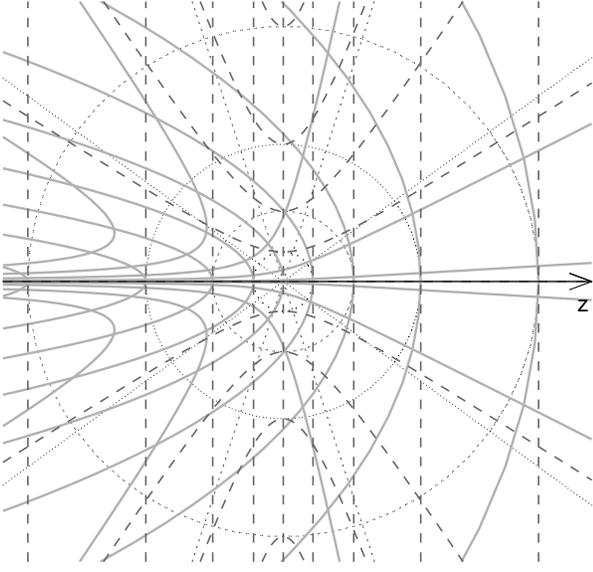}
 \caption{
 \figlabel{r-theta-rRW-thRW}
A section through the spherical, planar and pseudospherical
foliations of the $k = -1$ RW model.  The curves of constant $r_S$ and constant
$\theta_S$ are the thin dotted lines, the curves of constant $r_P$ are the solid
pale arches, the curves of constant $\theta_P$ are the solid pale lines
diverging from the left, the curves of constant $r_H$ are the medium dashed
vertical lines, and the curves of constant $\theta_H$ are the medium dashed
curves in the  left-right direction. This diagram is distorted because a
negatively curved 2-surface has been compressed onto a flat plane.  In fact all
three sets of lines are orthogonal. The 3-d diagram is obtained by rotating this
one about the $z$ axis.  At the right of the diagram it is evident how the 3
coordinate systems are similar near the z axis.  This becomes exact as $z \to
\infty$.
 }
 \end{figure}

To verify that $R$ cannot go to zero in pseudospherical
foliations, we write the RW metric in the form
 \begin{align}
   ds^2 = - dt^2 + {\cal S}^2(t) \big\{ x^2(r) \, dr^2
   + y^2(r) (d\theta^2 + z^2(\theta) \, d\phi^2) \big\} ~,
   \eqlabel{ds2RWxyz}
 \end{align}
and require that it satisfy the EFEs with the usual Friedmann equation
(\ref{SdotSq}) for $\cal S$. We find
 \begin{align}
   x = \frac{y'}{\sqrt{\alpha - k y^2(r)}\;}
   ~,~~~~~~~~
   \tdn{2}{z}{\theta} + \alpha \, z = 0
 \end{align}
 where $\alpha$ is arbitrary and $y(r)$ is not fixed, except when $\alpha = 0$
and $k = 0$, in which case
 \begin{align}
   \td{y}{r} = 0
   ~,~~~~~~~~
   \tdn{2}{z}{\theta} = 0
 \end{align}
 and $x(r)$ is not fixed.  Clearly, if $k = +1$ $\alpha$ must be positive, so
$z(\theta)$ must be a trig function, but if $k = -1$ $\alpha$ can have either
sign, so $z(\theta)$ may be a trig function or a hyperbolic trig function.  Thus
$y(r) = \cosh(r)$ gives (\ref{RWmetric-k-1eps-1}) above.  Notice however that if
$k = -1$ and $y(r)$ goes to zero somewhere, such as $y(r) = \sinh(r)$, then
$\alpha$ cannot be negative.

 \subsection{Matching the Szekeres Metrics to Vacuum}
 \seclabel{VacMatch}

We will now match the general planar and pseudospherical
Szekeres solutions to vacuum metrics. Although the vacuum metrics are very
different in each case, the matching can be solved at one go for all $\epsilon$.
Inspired by Bonnor's result \cite{Bonn1976a, Bonn1976b} that the quasi-spherical
Szekeres metric in its full generality can be matched to the Schwarzschild
solution, we will verify that the two other Szekeres solutions, in their full
generality, can be matched to the corresponding plane- or pseudospherically
symmetric vacuum solutions, respectively. The planar and
pseudospherical analogues of the Schwarzschild solution are
known, even if not well-known (\cite{CaDe1968}, eq. 13.48 in Ref.
\cite{SKMHH03}). They can be written in one formula as
\begin{align}
{\rm d} s^2 & = - \left(\epsilon - \frac {2m} R\right) {\rm d} T^2 + \frac 1
{\epsilon - 2m / R}\ {\rm d} R^2 \nn \\
&~~~~+ R^2 \left[{\rm d} \vartheta^2 + \frac 1 {\epsilon}\ \sin^2
\left(\sqrt{\epsilon} \vartheta\right) {\rm d} \varphi^2\right], \label{7.23}
\end{align}
where $m$ is a constant and $\epsilon = \pm 1, 0$. The metric with $\epsilon =
1$ is the Schwarzschild solution; with $\epsilon = 0$ and $m > 0$ it is the
Kasner solution in untypical coordinates, as is easy to verify%
 \footnote{The transformation to the well-known form is
 $R = (9 t^2 m/2)^{1/3}$,~
 $T = (3/(4 m))^{1/3} z$,~
 $\vartheta = (2/(9 m))^{1/3} \sqrt{x^2 + y^2}\;$,~
 $\varphi = \tan^{-1}(x/y)$.
 }%
. With $\epsilon = -1$ we obtain the vacuum pseudospherically
symmetric metric.

Note that the vacuum metrics with $\epsilon \leq 0$ are very different from
Schwarzschild's. The Schwarzschild metric is static for $R > 2m$ and nonstatic
(vacuum Kantowski-Sachs) for $R < 2m$, but the two regions together form one
complete manifold, as evidenced by the Kruskal-Szekeres extension. The two other
metrics are {\it globally} nonstatic when $m > 0$, as $T$ is a space coordinate
and $R$ is time.
The Kasner solution with $m > 0$ is of Bianchi type I, the pseudospherical
one is of Bianchi type III (see Appendix \ref{Bianchi}).

First, we write the vacuum metrics in Szekeres form.  Following the prescription
used for the Schwarzschild solution (see Exercise 10 in Chap. 14 in Ref.
\cite{PlKr2006}), we transform (\ref{7.23}) to coordinates defined by observers
freely falling in the $R$-direction --- the plane- and pseudospherically
symmetric analogues of the Lema\^{\i}tre-Novikov coordinates
for the Schwarzschild solution, see \cite{Hel87} and Sec. 14.12 in Ref.
\cite{PlKr2006}. We substitute
\begin{equation}\label{7.24}
T = T(t, r), \qquad R = R(t, r)
\end{equation}
and require that in the $(t, r)$ coordinates the component $g_{tt}$ of the
metric is $-1$, while $g_{tr} = 0$. We solve this set of equations for $T_{,t}$
and $T_{,r}$, then impose the integrability condition $T_{,tr} = T_{,rt}$.
Discarding the trivial case $R_{,r} = 0$, it reduces to
\begin{equation}\label{7.25}
{R_{,t}}^2 = \frac {2m} R + F(r),
\end{equation}
where $F(r)$ is an arbitrary function. This is a special case of eq.
(\ref{2.5}), corresponding to $M = m =$ const and $F = f$. The full solution for
$T(t,r)$ is given in \cite{Hel96b}. The metric (\ref{7.23}) in the $(t, r)$
coordinates becomes
\begin{equation}\label{7.26}
{\rm d} s^2 = - {\rm d} t^2 + \frac {{R,_r}^2} {\epsilon + F}\ {\rm d} r^2 + R^2
\left[{\rm d} \vartheta^2 + \frac 1 {\epsilon}\ \sin^2 \left(\sqrt{\epsilon}
\vartheta\right) {\rm d} \varphi^2\right],
\end{equation}
and the coordinates of (\ref{7.26}) are adapted to matching it to the planar or
pseudospherical Szekeres metrics across a hypersurface of constant $r$.

The matching requires that the intrinsic metric of a hypersurface $r =$ const
and the second fundamental form of this hypersurface are the same for both
4-metrics. The match between the 3-metrics follows easily. Suppose the matching
is done at $r = b$. The transformations to be applied to (\ref{7.26}) are
different for each value of $\epsilon$.  With $\epsilon = 0$ we
transform the coordinates of (\ref{7.26}) as follows
\begin{eqnarray}\label{7.27}
(\vartheta, \varphi) &=& \left(2 \sqrt{{p'}^2 + {q'}^2}, \arctan
(q'/p')\right), \nonumber \\
(p', q') &=& \frac {S(b) \left(p - P(b), q - Q(b)\right)} {(p - P(b))^2 + (q -
Q(b))^2}.
\end{eqnarray}
With $\epsilon = -1$, we transform (\ref{7.26}) by
\begin{align}\label{7.28}
\tanh (\vartheta / 2) &= \frac 1 {S(b)}\ \sqrt{(p - P(b))^2 + (q - Q(b))^2},
\nonumber \\
\varphi &= \arctan \left[\frac {q - Q(b)} {p - P(b)}\right].
\end{align}
After the transformation (\ref{7.27}), or, respectively, (\ref{7.28}), the
metric (\ref{7.26}) becomes
\begin{align}\label{7.29}
{\rm d} s^2 &= - {\rm d} t^2 + \frac {{R,_r}^2} {\epsilon + F}\ {\rm d} r^2 +
\frac {R^2} {{E_1}^2} \left({\rm d} p^2 + {\rm d} q^2\right), \nonumber \\
E_1 &= \frac {S(b)} 2\left[\frac {\left(p - P(b)\right)^2 + (q - Q(b))^2}
{(S(b))^2} + \epsilon\right].
\end{align}
In the single $r = b$ hypersurface the 3-metric of (\ref{7.29}) has the same
form as in (\ref{2.1}). The two 3-metrics will coincide if their $R(t, b)$ are
the same at all times. This will be the case when
\begin{equation}\label{7.30}
M(b) = m ~, \qquad f(b) = F(b) ~,
\end{equation}
since then both $R$-s obey the same differential equation, so it is enough to
choose the same initial condition for both of them. The unit normal vector to
the matching hypersurface, $n_\alpha$ in the Szekeres metric is
\begin{equation}\label{7.31}
   n_{S \alpha} = (0, n_1, 0, 0) ~, \qquad
   n_1 = \frac {R' - RE'/E} {\sqrt{\epsilon + f}} ~,
\end{equation}
and in the vacuum metric (\ref{7.29}) it is $n_{e \alpha} = (0, R' /
\sqrt{\epsilon + f}, 0, 0)$. In spite of these different forms, the terms $(R' -
RE'/E)$ and $R'$ cancel out in the extrinsic curvature for each metric, and the
only nonvanishing components of the second fundamental form of the $r = b$
hypersurface $K_{ij}$ are $K_{22} = K_{33} = - \sqrt{\epsilon + f} R / E^2$
which are continuous across $r = b$ by virtue of (\ref{7.30}) and (\ref{7.29}).

We see that, with the Szekeres mass function $M$ being positive, the matching
implies $m > 0$ in both cases, and so the ``exterior" vacuum solutions for
$\epsilon \leq 0$ are {\it necessarily nonstatic}. This, in turn, implies that
any Szekeres dust model that matches on to them cannot be in a static state.

Thus, in the most general case, the exterior metric for the planar Szekeres
metric is the vacuum Kasner metric, and for the pseudospherical
Szekeres metric it is the $\epsilon = -1$ vacuum metric (\ref{7.23}), both
represented as in (\ref{7.29}).

 \section{A flat model of the Szekeres spaces with $\epsilon \neq0$}
 \seclabel{flatmodel}

To visualise the geometric relations in the Szekeres spatial sections, we will
construct an analogue of the Szekeres coordinate system in a flat 3-space.

 \subsection{The quasi-spherical case.}

For the beginning, we will deal with the $\epsilon = +1$ case, i.e. with the
foliation by nonconcentric spheres. We first construct the appropriate
coordinates in a plane. The setup will be axially symmetric, and after the
construction is completed we will add the third dimension by rotating the
whole set around the symmetry axis. The foliating spheres intersect the plane along
nonconcentric circles. The family of circles is such that their radii increase
from $0$ to $\infty$ while the positions of their centers move from the point
$(b, 0)$ to $(+ \infty, 0)$ in such a way that in the limit of infinite radius
the circles tend to the vertical line $x = 0$.

The family of circles is shown in the right half of Fig. \ref{fullszekeres}; it is
given by the equation:
\begin{equation}\label{8.1}
\left(x - \sqrt{b^2 + u^2}\right)^2 + y^2 = u^2 ~, \\
\end{equation}
where $b$ is a constant that determines the position of the center of the
limiting circle of zero radius which we will call the origin $O$, while $u$ is
the parameter of the family -- the radius of the circles.

For this family, we now construct a family of orthogonal curves. The tangents to
the family (\ref{8.1}) have slope
\begin{equation}\label{8.2}
\dr y x = \frac {y^2 - x^2 + b^2} {2 x y}.
\end{equation}
which is the differential equation whose solution is (\ref{8.1}). The orthogonal
curves will obey the equation
\begin{equation}\label{8.3}
\dr y x = \frac {-2 x y} {y^2 - x^2 + b^2} ~\Longleftrightarrow~ \dr x y = \frac
{x^2 - y^2 - b^2} {2 x y}.
\end{equation}
Note that this results from (\ref{8.2}) by the substitution:
\begin{equation}\label{8.4}
(x, y, b) = (y', x', {\rm i} b').
\end{equation}
Thus a solution of (\ref{8.4}) results from (\ref{8.1}) by the same substitution
and it is:
\begin{equation}\label{8.5}
x^2 + \left(e_1 y - \sqrt{v^2 - b^2}\right)^2 = v^2,
\end{equation}
where $e_1 = \pm 1$ and $v$ is the parameter of the orthogonal family. As it
happens, (\ref{8.5}) is also a family of nonconcentric circles whose centres all
lie on the $y$-axis, but the radius of the smallest circle is $b$. All the
circles pass through the origin $O$ at $(x, y) = (b, 0)$. Fig.
\ref{fullszekeres} shows the $x > 0$ part of both families. The double sign in
(\ref{8.5}) is needed to cover the whole right half of Fig. \ref{fullszekeres}.
With only the $+$ sign, only the $y > 0$ sector would be covered. We did not
include the corresponding double sign in (\ref{8.1}) because we wanted to cover
only the $x
> 0$ half-plane with those circles.

We now choose $u$ and $v$ as the coordinates on the plane and calculate the
metric in these coordinates. From (\ref{8.1}) and (\ref{8.5}) we find
\begin{align}
& x = \frac {b^2 v} {D}, \qquad y = e_1 e_2 \frac {b^2 u} {D}, \nn \\
& D \df v \sqrt{u^2 + b^2} + e_2 u \sqrt{v^2 - b^2} ~, \label{8.6}
\end{align}
where $e_2 = \pm 1$. The two solutions arise because, as seen from Fig.
\ref{fullszekeres}, the pair of circles corresponding to a given pair of values
of $(u, v)$ in general intersects in two points. The exceptional cases are $u =
0$ (which corresponds to the single point $(x, y) = (b, 0)$) and $v = \pm b$ --
when the $v$ circle is mirror-symmetric in $y$, and the two intersection points
have the same $x$ coordinate. Using (\ref{8.6}) we find
\begin{eqnarray}\label{8.7}
{\rm d} x^2 + {\rm d} y^2 &=& \frac {b^4} {D^2}\ \left(\frac {v^2} {u^2 + b^2} \
{\rm d} u^2 + \frac {u^2} {v^2 - b^2}\ {\rm d} v^2\right).\ \ \
\end{eqnarray}

To make the metric look more like Szekeres, we now transform the coordinate
$v$ as follows:\footnote{The transformation is a composition of two
transformations: $v = b / \sin \chi$ and $\chi = 2 \arctan [(w/(2b)]$.}
\begin{equation}\label{8.8}
v = \frac {b^2 + w^2/4} w ~,
\end{equation}
after which the metric (\ref{8.7}) becomes
\begin{align}\label{8.9}
&{\rm d} x^2 + {\rm d} y^2 = \frac 1 {\widetilde{E}^2}\ \left[\frac {\left(b^2 +
w^2/4\right)^2} {u^2 + b^2}\ {\rm d} u^2 + u^2 {\rm d} w^2\right], \nonumber \\
& \widetilde{E} = w D / b^2 \nonumber \\
&= \sqrt{u^2 + b^2} - u + \left(\sqrt{u^2 + b^2} + u\right) \frac {w^2} {4 b^2}.
\end{align}
In the above, for a more explicit correspondence with the Szekeres solution, we
have chosen $e_2 = -1$ and $\sqrt{\left(b^2 - w^2/4\right)^2} = + \left(b^2 -
w^2/4\right)$, so that the term independent of $w$ tends to zero as $u \to
\infty$. This will correspond to $\epsilon \to 0$ in the Szekeres metric. (The
case $e_2 = +1$ results from (\ref{8.9}) by the inversion $w = 4b^2/w'$.)

By looking at the Szekeres metric (\ref{2.1}) we see that in (\ref{8.9}) $u$
simultaneously plays the role of $r$ and of $R$. Let us follow the analogy. We
are considering a flat 3-space (so far, only 2-plane). The 3-space $t =$ const
in the $\epsilon = +1$ Szekeres metric will be flat when $f = 0$. Thus, if
(\ref{8.9}) is to become the metric of a flat space $t =$ const in the $\epsilon
= +1$ Szekeres metric, then the coefficient of ${\rm d} u^2$ in (\ref{8.9})
should obey:
\begin{equation}\label{8.10}
\frac {\left(b^2 + w^2 / 4\right)^2} {\left(u^2 + b^2\right) \widetilde{E}^2} =
\left[\frac 1 {\widetilde{E}}\ \left(\widetilde{E} - u
\widetilde{E},_u\right)\right]^2.
\end{equation}
As can be verified, this holds.

Now it remains to add the third dimension by rotating the whole configuration
around the $x$ axis of the initial coordinates. Thus, in $\left({\rm d} x^2 +
{\rm d} y^2\right)$ we now treat $y$ as a radial coordinate, we add $\phi$ as
the angle of the polar coordinates, and consider the metric $\left({\rm d} x^2 +
{\rm d} y^2 + y^2 {\rm d} \phi^2\right)$. We go back to (\ref{8.6}) and repeat
the calculations with this 3-dimensional metric.
Thus, going to the Cartesian coordinates $(\widetilde{y}, z) = (y \cos \phi, y
\sin \phi)$ we thereby transform $(w, \phi)$ to $(p, q) = w (\cos \phi, \sin
\phi)$, or
\begin{equation}\label{8.11}
w = \sqrt{p^2 + q^2}, \qquad \phi = \arctan (q/p),
\end{equation}
and after this the metric becomes
\begin{eqnarray}\label{8.12}
&&{\rm d} {s_3}^2 = \left[1 - \frac {u E_{,u}} E\right]^2\ {\rm d}u^2 +
\frac {u^2} {E^2} \left({\rm d} p^2 + {\rm d} q^2\right) ~, \nn \\
&&E = \sqrt{u^2 + b^2} - u + \frac {\sqrt{u^2 + b^2} + u} {4 b^2} \left(p^2 +
q^2\right).\ \ \ \ \ \
\end{eqnarray}
This is the axially symmetric (and flat) subcase of the 3-space $t =$ const in
the Szekeres metric of (\ref{2.1}) -- (\ref{2.5}) corresponding to $\epsilon =
+1$, $B_1 = B_2 = P = Q = 0$, $S = 2b^2 / \left(\sqrt{u^2 + b^2} + u\right)
\equiv 2 \left(\sqrt{u^2 + b^2} - u\right)$ and $r = R = u$.

 \begin{figure}
 \begin{center}
 \includegraphics[scale=0.4]{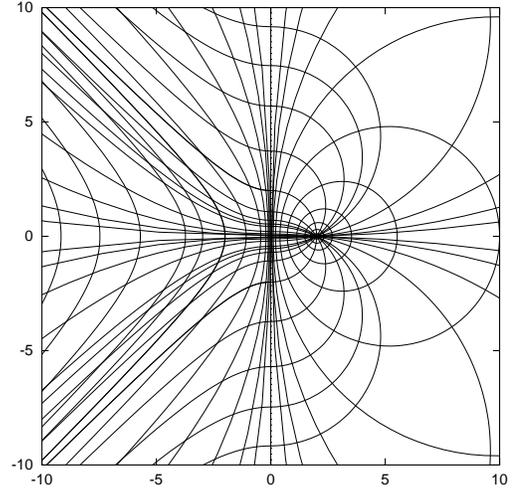}
 \caption{
 \figlabel{fullszekeres}
 \footnotesize
A section through the family of nonconcentric spheres going over into a family
of hyperboloids via a plane, that provides a model of flat space sections in
Szekeres coordinates. The curves orthogonal to them are also shown.  More
explanation in the text. }
 \end{center}
 \end{figure}

\subsection{The Pseudospherical Case}

We can deal with the quasi-pseudospherical case in a similar
way, but there is an important difference; in this case the 3-space of constant
time can be flat only if its metric is pseudoeuclidean. This pseudoeuclidean
space represents a space of Euclidean signature that has constant negative
curvature. We will plot the coordinate curves in a Euclidean space, so it has to
be remembered that they are distorted and do not represent the geometric
relations faithfully. In particular, vectors or curves that are orthogonal in
the pseudoeuclidean metric will not look orthogonal in the plot. The space $t =$
const can have a Euclidean signature, but then it must be curved. So we can say
that we are representing the relations in a curved space with $f \neq 0$ by
figures drawn on a flat plane, and hence the distortion.

We begin with a family of right hyperbolae that fill the left half of the $(x,
y)$ plane and in the limit $u \to \infty$ tend to the straight line $x = 0$; at
the end of the calculation we will rotate the whole collection around the
$x$-axis. The family is given by
\begin{equation}\label{8.13}
\left(x - \epsilon_1 \sqrt{u^2 - b^2}\right)^2 - y^2 = u^2,
\end{equation}
where $\epsilon_1 = \pm 1$%
\footnote{This double sign is necessary in order that the hyperbolae fill the
whole left half-plane. With only one sign they would fill only the part of the
half-plane that lies to one side of the $u = b$ curve.} and $u \geq b$ is the
parameter of the family. The family is shown in the left part of Fig.
\ref{fullszekeres}.

We again construct the family of curves orthogonal to these hyperbolae, but in
the pseudoeuclidean sense. The hyperbolae (\ref{8.13}) solve the differential
equation
\begin{equation}\label{8.14}
\dr y x = \frac {x^2 + y^2 + b^2} {2 x y},
\end{equation}
so the curves that are (pseudo) orthogonal to them obey the
equation:
\begin{equation}\label{8.15}
\dr y x = \frac {2 x y} {x^2 + y^2 + b^2},
\end{equation}
which is obtained from (\ref{8.14}) simply by interchanging $x$ and $y$. We thus
conclude that the solution is obtained from (\ref{8.13}) by the same
interchange, and so it is
\begin{equation}\label{8.16}
\left(y - \epsilon_2 \sqrt{v^2 - b^2}\right)^2 - x^2 = v^2,
\end{equation}
where $\epsilon_2 = \pm 1$. The two families (\ref{8.13}) and (\ref{8.15}) are
shown together in the left half of Fig. \ref{fullszekeres}.

By solving (\ref{8.13}) and (\ref{8.16}) for $x$ and $y$ we find:
\begin{align}
& x = - \frac {b^2 v} {D_h}, \qquad y = \epsilon_2 \epsilon_3 \frac {b^2 u}
{D_h}, \nn \\
& D_h \df \epsilon_1 v \sqrt{u^2 - b^2} + \epsilon_3 u \sqrt{v^2 - b^2},
\label{8.17}
\end{align}
where $\epsilon_1$, $\epsilon_2$ $\epsilon_3 = \pm 1$; from which we get
\begin{equation}\label{8.18}
- {\rm d} x^2 + {\rm d} y^2 = \frac {b^4} {{D_h}^2} \left(- \frac {v^2 {\rm d}
u^2} {u^2 - b^2} + \frac {u^2 {\rm d} v^2} {v^2 - b^2}\right).
\end{equation}
Again substituting for $v$ with (\ref{8.8}), and adding the 3rd dimension by
rotating around the $x$ axis in a similar way to (\ref{8.11}), the three
dimensional metric $\left(- {\rm d} x^2 + {\rm d} y^2 + y^2 {\rm d}
\varphi^2\right)$ becomes
\begin{align}\label{8.19}
{\rm d} {s_3}^2 &= \frac 1 {{E_1}^2}\ \left\{\frac {- \left[b^2 + \frac 1 4
\left(p^2 + q^2\right)\right]^2} {u^2 - b^2}\ {\rm d} u^2\right. \nonumber \\
&+ \left.u^2 \left({\rm d} p^2 + {\rm d} q^2\right)\right\}, \nonumber \\
E_1 &= \sqrt{u^2 - b^2} - u + \frac {\sqrt{u^2 - b^2} + u} {4 b^2} \left(p^2 +
q^2\right).
\end{align}
Just as for the spherical case, it can be verified that the Szekeres relation,
analogous to (\ref{8.10}), is obeyed. It reads here
\begin{equation}\label{8.20}
\frac {\left[b^2 + \frac 1 4 \left(p^2 + q^2\right)\right]^2} {\left(u^2 -
b^2\right) {E_1}^2} = \frac {\left(E_1 - u E_{1,u}\right)^2} {{E_1}^2}.
\end{equation}
For the same reason as with (\ref{8.9}), we have now chosen $\epsilon_3= -
\epsilon_1$ and $\sqrt{\left(b^2 - w^2/4\right)^2} = + \left(b^2 -
w^2/4\right)$. Then the sign of $\epsilon_1$ becomes irrelevant, since $D_h
\propto \epsilon_1$, and only ${D_h}^2$ appears in the metric. The metric
(\ref{8.20}) corresponds to (\ref{2.1}) with $\epsilon = -1$, $B_1 = B_2 = P = Q
= 0$, $S = 2b^2 / \left(\sqrt{u^2 - b^2} + u\right) \equiv 2 \left(\sqrt{u^2 -
b^2} - u\right)$ and $r = R = u$.

Fig. \ref{fullszekeres} shows the junction of the spaces of (\ref{8.12}) and
(\ref{8.19}) -- it represents the Szekeres $t =$ const space consisting of
nonconcentric spheres (right half of the picture) that tend to the plane $x = 0$
from one side, and the family of hyperboloids (left half of the picture) that
tend to the same plane from the other side. This shows how spherical surfaces
can go over into hyperboloidal surfaces within the same space. We repeat that
only the right half of the picture faithfully represents the geometry of the
flat space in coordinates defined by the spheres; the left half is a distorted
image of either a curved 3-space or of a flat 3-space that has the
pseudoeuclidean signature $(- + +)$.

Note that the plane that separates the family of spheres from the family of
hyperboloids has, in each family, the equation $u \to \infty$ (to see this,
solve (\ref{8.1}) and (\ref{8.13}) for $x$; in each case one of the solutions
resulting when $u \to \infty$ is the plane $x = 0$). In this limit $\sqrt{u^2
\pm b^2} - u \to 0$, which, on comparing (\ref{8.12}) and (\ref{8.19}) with
(\ref{Edef}), is seen to correspond to $\epsilon = 0$, just as it should.

 \section{Physical Discussion}

 \subsection{Role of $R$}

In the metric (\ref{2.1}) and in the area integral, $A = R^2 \int 1/E^2 \, dp \,
dq$, the factor $R^2$ multiplies the unit sphere or pseudosphere, and therefore
determines the magnitude of the curvature of the constant $(t,r)$ surfaces.  It
is also a major factor in the curvature of the constant $t$ 3-spaces. Therefore
we view it as an ``areal factor" or a ``curvature scale". However, when
$\epsilon \leq 0$ it is not at all like a spherical radius.  We note that when
$\epsilon = -1$, there can be no origin, but $R$ can have maxima and minima as
$r$ varies, while in the $\epsilon = 0$ case, $R$ cannot have extrema, and it
can only approach zero asymptotically.

 \subsection{Role of $M$}

In (\ref{2.5}), $M$ looks like a mass in the gravitational potential energy term
of the evolution equation (\ref{2.5}), while in (\ref{Rddot}) $M$ determines the
deceleration of $R$.  For $\epsilon = +1$, the function $M(r)$ plays the role of
the gravitational mass contained within a comoving ``radius'' $r$, but this
interpretation is geometrically and physically correct only in the
quasi-spherical model, where the surfaces of constant $r$ are nonconcentric
spheres enclosing a finite amount of matter.  For $\epsilon \leq 0$ however, $R$
is not the spherical radius that is an important part of these ideas in their
original form, and $M$ is not a total (gravitational) mass, since the constant
$t$ \& $r$ surfaces are not closed.  Consequently these ideas need revising.

The impossibility of an ``origin" or locus where $M$ and $R$ go to zero when
$\epsilon = -1$ means that $M$ must have a global minimum, and indeed regular maxima
and minima in $R$ and $M$ are possible.  Therefore the local $M$ value is not
independent of it's value elsewhere, and integrals of the density over a region
always have a boundary term, suggesting the value of $M$ (rather than its change
between two shells) is more than can associated with any finite part of the mass
distribution.  In $\epsilon = 0$ models, an asymptotic ``origin" is possible, but
not required, and regular maxima and minima in $R$ and $M$ are also possible
asymptotically.  So, with an asymptotic origin (as occurs in the planar foliation of
$k = -1$ RW) the boundary term can be set to zero, but not with an asymptotic minimum
in $M$ and $R$.

Nevertheless, the central roles of $R$ and $M$ are confirmed by the fact that
the 3 types of Szekeres model can be joined smoothly to vacuum across a constant
$r$ surface at which the values of $R$ and $M$ must match (section
\ref{VacMatch}). The vacuum metric ``generated" by the Szekeres dust
distribution must have spherical, planar, or pseudospherical
symmetry, and in each $M$ is the sole parameter, while $R$ is an areal factor.

We note that, even in the Poisson equation, the gravitational potential does not
need to be associated with a particular body of matter, and indeed it is not
uniquely defined for a given density distribution.

Therefore we find that $M$ is the mass factor in the gravitational potential
energy.

 \subsection{Role of $f$}

As shown in section \ref{3spaces}, and as is apparent from the metric
(\ref{2.1}), the function $f$ determines sign of the curvature of the 3-space $t
=$~const, as well as being a factor in its magnitude (c.f. \cite{Hel87}).  In
the quasi-spherical case, $\epsilon = +1$, this 3-space becomes Euclidean
(represented in odd coordinates) when $f = 0$.  In the quasi-pseudospherical
case, with $f = 0$ it becomes flat but pseudoeuclidean: the
signature is $(- + +)$.  In the quasi-planar case, the equations show the value
$f = 0$ is not possible, and thus the quasi-planar case does not in fact admit
flat 3-dimensional subspaces.

We also see from above that $f$ appears in the gravitational energy equation as
the total energy per unit mass of the matter particles, and we do not need to
revise this interpretation.  Therefore, this variable has the same role as in
quasi-spherical and spherically symmetric models.

 \subsection{Role of $E$}

We have seen in section \ref{RoleE} that for $\epsilon = +1$, $E'/E$ is the
factor that determines the dipole nature of the constant $r$ shells, and for
$\epsilon = -1$, it is the pseudospherical equivalent of a
dipole, except that the two sheets of the hyperboloid contain half the dipole
each, and only one of them can be free of shell crossings.  For $\epsilon = 0$,
the effect of $E'/E$ is merely to tilt adjacent shells relative to each other,
but only the zero tilt case ($E' = 0$) is free of shell crossings.

The shell separation (along the $r$ lines) decreases monotonically as $E'/E$
increases.  If $E' = 0$ it is uniform, otherwise it is minimum at some location
and diverges outwards.

For pseudospherical models, which must have $f \geq 1$, (\ref{R'RM'3Mpos}) and
(\ref{rholim}) show that if $f'/(2f) \geq M'/(3M)$ everywhere and there are no shell
crossings, the density is at all times monotonically decreasing with $E'/E$, but
asymptotically approaches a finite value as $E'/E$ diverges.  Therefore the density
distribution on each shell is that of a void, but the void centres on successive
shells can be at different $(p, q)$ or $(\theta, \phi)$ positions, in other words,
the void has a snake-like or wiggly cylinder shape. The minimum density is only
zero if $M'/(3M) = -(E'/E)_\text{max}$.  Far from the void, at large $\theta$, the
density is asymptotically uniform with $p$ \& $q$ (i.e. with $\phi$), but can vary
with $r$, though fairly gently compared with the void interior.  If
$f'/(2f) < M'/(3M)$ everywhere , an initial void can evolve into an overdensity.
Intuitively, it makes sense that there should be an initial underdensity,
since too strong a tube-like overdensity would cause outer shells to expand much
less rapidly, but this would cause shell crossings in models with hyperbolic
evolution, $f > 0$.

The location of the density minimum (or maximum) on each sheet is given by
(\ref{TanPhiX}) and (\ref{TanhThetaX}), so their values are limited by the no
shell crossings condition (\ref{maxE'/Eexists}).  Their rates of change depend
on $S''$, $P''$ and $Q''$ and since the latter are not directly limited, they
could be arbitrarily large at any one point, however (\ref{maxE'/Eexists})
implies that for any given $r_1$ and $r_2$
 \begin{align}
   2 \int_{r_1}^{r_2} (P' P'' + Q' Q'') \, dr + (P'_1)^2 + (Q'_1)^2 \nn \\
   \leq 2 \int_{r_1}^{r_2} S' S'' \, dr + (S'_1)^2
 \end{align}
which means there is a limit on how far the location of the minimum can move for
a finite change in $r$.
The density is affected by $E'/E$ at all times except near a simultaneous bang
or crunch.

 \section{Conclusions}

We have analysed the Szekeres metrics with quasi-pseudospherical
and quasi-planar spatial foliations, and established their regularity
conditions and their physical properties.

For the quasi-pseudospherical case ($\epsilon = -1$), each constant $r$ shell is
a two-sheeted right hyperboloid (pseudosphere), each mapping to only part of the
$p$-$q$ plane, but only one sheet can be free of shell crossings, and only if
$E'/E$ has a negative maximum, going to $-\infty$ in the asymptotic regions of
the sheet (where $E \to 0$). The effect of $E'/E$ can be called the
pseudospherical equivalent of a dipole, but half the dipole is in the disallowed
sheet. At this maximum the constant $r$ shells are closest and the density has
an extremum --- a minimum if (\ref{R'RM'3Mpos}) holds, otherwise it starts as a
minimum, but evolves into a maximum.  Far from the the extremum, the density
becomes uniform with $p$ \& $q$, but can still vary with $r$. The location and
value of the extremum depend on the derivatives of $S(r)$, $P(r)$ and $Q(r)$,
and so the density extremum can vary in magnitude and makes a wiggling,
snake-like path. We also find that on a spatial section $R$ can have extrema,
but cannot be zero.  The conditions for no shell crossing are weaker than for LT
models, allowing $R'$ or $M'$ to become negative, though there is an extra
condition relating the Szekeres functions $S'$, $P'$ and $Q'$. (In contrast, for
spherical foliations the no shell crossing conditions are not weaker than in
LT.)

We found the quasi-planar case ($\epsilon = 0$) was the hardest to understand.
Only the plane symmetric case, $E' = 0$ can be free of shell crossings, spatial
sections can have neither zeros nor extrema of $R$, except asymptotically, and
it isn't possible to make it spatially flat, $f = 0$ (except in the
Kantowski-Sachs-like limit \cite{Hel96}),
so as a complete manifold it turns out to be the most
restricted once physical regularity conditions are imposed.
However a Szekeres spacetime can consist of a region with a quasi-spherical
foliation joined across a planar boundary to a region with a
quasi-pseudospherical foliation, as visualised in the the 3-d model of section
\ref{flatmodel}.

It was necessary to take particular care in analysing the meaning of $R$ and
$M$.  Although the evolution of $R$ obeys an energy equation with a term $M/R$
that is very like a gravitational potential, $R$ cannot be a spherical radius as
the surfaces it multiplies are not closed. Similarly, because there's always a
boundary term when $\epsilon = -1$, $M$ is not solely determined by the matter
inside a finite region, though its change in value between two constant $r$
shells may possibly be associated with the matter bewteen them.

Nevertheless, $R$ is very closely tied to the curvature of the $p$-$q$
2-surfaces and to their areas, so it is a ``curvature scale" or an ``areal
factor".  The Poisson equation and the EFEs only relate field derivatives to the
local matter, so in general the gravitational potential $\Phi$ has no simple
connection to a volume integral of the density $\rho$. Similarly, in the planar
and pseudospherical foliations studied here, we see an example in which the
gravitational potential energy term in the $R$ evolution equation (\ref{2.5}) is
affected not only by the density and curvature in a finite region, but also by
boundary values determined by the distant density and curvature distribution. We
view $M$ as a ``potential mass" since it is a quantity with dimension mass that
determines a gravitational potential energy through $M/R$, and acceleration
through $M/R^2$. $M$ is the key gravitational field parameter, and it relates
the curvature scale $R$ of the comoving surfaces to the potential energy of the
$\dot{R}$ equation.

Having understood some key features of the $\epsilon \leq 0$ Szekeres metrics,
it should now be easier to construct useful models out of them.

 \begin{acknowledgements}
The work of AK was partly supported by the Polish Ministry of Science and
Education grant no 1 P03B 075 29. AK expresses his gratitude for the
Department of Mathematics and Applied Mathematics in Cape Town, where most of
this work was done, for hospitality and perfect working conditions. CH thanks
the South African National Research Foundation for a grant. This research was
supported by an award from the Poland-South Africa Technical Agreement.
 \end{acknowledgements}

 \appendix

 \section{The $E' = 0$ Locus as a Geodesic of the 2-d Hyperboloid}
 \seclabel{Geod}

The set $E' = 0$ in the $(p, q)$ surface of the metric (\ref{2.1}) with
$\epsilon = -1$ is a geodesic in that surface. Proof:

Calculate $E'$ and rewrite the result in the $(\theta, \phi)$ variables of
(\ref{Riemprojm}):
\begin{equation}\label{A.1}
E' = - \frac {S' \cosh \theta + \sinh \theta (P' \cos \phi + Q' \sin \phi)}
{\cosh \theta - 1}.
\end{equation}
The solution of the equation $E' = 0$ is
\begin{equation}\label{A.2}
\tanh \theta = - \frac {S'} {P' \cos \phi + Q' \sin \phi}.
\end{equation}
Choose $\phi$ as the parameter on the curve $\theta(\phi)$ given by (\ref{A.2}).
The tangent vector to this curve then has the components
\begin{equation}\label{A.3}
k^{\alpha} = \left(\dr {\theta} {\phi}, 1\right), \quad \dr {\theta} {\phi} =
\frac {S' (- P' \sin \phi + Q' \cos \phi)} {(P' \cos \phi + Q' \sin \phi)^2 -
{S'}^2}.
\end{equation}
The metric (\ref{2.19}), and its nonzero Christoffel symbols, in the $(\theta,
\phi)$ coordinates, are
\begin{eqnarray}\label{A.4}
{\rm d} {s_2}^2 &=& {\rm d} \theta^2 + \sinh^2 \theta {\rm d} \phi^2,
\nonumber \\
\Chr{1\ } {2} {2} &=& - \sinh \theta \cosh \theta, \qquad \Chr{2\ } {1} {2} =
\coth \theta.
\end{eqnarray}
Thus the equations of a geodesic are
\begin{eqnarray}\label{A.5}
\dr {^2 \theta} {\phi^2} - \sinh \theta \cosh \theta &=& \lambda \dr {\theta}
{\phi}, \nonumber \\
2 \coth \theta \dr {\theta} {\phi} &=& \lambda,
\end{eqnarray}
where $\lambda$ is an unknown proportionality factor. The second equation
above defines $\lambda$, which is
\begin{equation}\label{A.6}
\lambda = - \frac {2 (- P' \sin \phi + Q' \cos \phi) (P' \cos \phi + Q' \sin
\phi)} {(P' \cos \phi + Q' \sin \phi)^2 - {S'}^2},
\end{equation}
and then the first of (\ref{A.5}) is easily verified using (\ref{A.3}) and
\begin{align}
\sinh \theta \cosh \theta & = \frac {\tanh \theta} {1 - \tanh^2 \theta} \nn \\
& = - \frac {S' (P' \cos \phi + Q' \sin \phi)} {(P' \cos \phi + Q' \sin \phi)^2 -
{S'}^2}
\end{align}
$\square$.

\section{The Bianchi type of the pseudospherical vacuum
model.}\seclabel{Bianchi}

Lower indices will label vectors, the upper indices will label the coordinate
components of vectors. The Killing vector fields for the metric (\ref{7.23})
with $\epsilon = -1$ are:
\begin{eqnarray}\label{C.1}
{k_1}^{\alpha} &=& {\delta_1}^{\alpha}, \qquad {k_4}^{\alpha} =
{\delta_3}^{\alpha}, \nonumber \\
{k_2}^{\alpha} &=& \cos \varphi {\delta_2}^{\alpha} - \coth \vartheta \sin
\varphi {\delta_3}^{\alpha}, \nonumber \\
{k_3}^{\alpha} &=& \sin \varphi {\delta_2}^{\alpha} + \coth \vartheta \cos
\varphi {\delta_3}^{\alpha}.
\end{eqnarray}
The commutators are:
\begin{eqnarray}\label{C.2}
\left[k_1, k_I\right] &=& 0, \qquad I = 1, 2, 3, \nonumber \\
\left[k_2, k_3\right] &=& k_4, \qquad \left[k_2, k_4\right] = k_3, \nonumber \\
\left[k_3, k_4\right] &=& - k_2.
\end{eqnarray}
The Bianchi algebra must thus include $k_1$ and a 2-dimensional subspace of
$\left\{k_2, k_3, k_4\right\}$. Consequently, out of the set $\left\{k_2, k_3,
k_4\right\}$ we choose two linear combinations, $\ell$ and $m$, that span a
2-dimensional Lie algebra, i.e. have the property $[\ell, m] = \alpha \ell +
\beta m$. This can be done in many ways; one example of such a combination is
\begin{equation}\label{C.3}
\ell = k_2, \qquad m = k_2 + k_3 + k_4,
\end{equation}
for which we have
\begin{equation}\label{C.4}
[\ell, m] = m - \ell.
\end{equation}
This is not a standard Bianchi basis. To obtain a standard basis (see Ref.
\cite{PlKr2006}) we take such combinations of $k_1$, $\ell$ and $m$ that are
equivalent to
\begin{eqnarray}\label{C.5}
w_1 &=& 2k_2 + k_3 + k_4, \qquad w_2 = k_1 - k_3 - k_4, \nonumber \\
w_3 &=& k_1 + k_3 + k_4.
\end{eqnarray}
The commutation relations are now
\begin{equation}\label{C.6}
\left[w_1, w_2\right] = w_2 - w_3 = \left[w_3, w_1\right], \qquad \left[w_2,
w_3\right] = 0,
\end{equation}
and this is the standard form of the Bianchi type III algebra.

 \end{document}